\documentclass[
 reprint,
superscriptaddress,
amsmath,amssymb,
aps,
floatfix,
]{revtex4-2}

\usepackage{graphicx}
\usepackage{dcolumn}
\usepackage{bm}
\usepackage[colorlinks=true,allcolors=blue]{hyperref}
\newcommand{\ee}{\textrm{e}}
\newcommand{\mm}{\textrm{m}}
\newcommand{\bb}{\textrm{b}}

\usepackage{siunitx}
\begin{document}

\title{Mechanistic rules for de novo design of enzymes}

\author{Michalis Chatzittofi}
\address{Max Planck Institute for Dynamics and Self-Organization (MPI-DS), D-37077 Göttingen, Germany} 
\author{Jaime Agudo-Canalejo}
\email{j.agudo-canalejo@ucl.ac.uk}
\address{Max Planck Institute for Dynamics and Self-Organization (MPI-DS), D-37077 Göttingen, Germany} 
\address{Department of Physics and Astronomy, University College London, London WC1E 6BT, United Kingdom}
\author{Ramin Golestanian}
\email{ramin.golestanian@ds.mpg.de}
\address{Max Planck Institute for Dynamics and Self-Organization (MPI-DS), D-37077 Göttingen, Germany} 
\address{Rudolf Peierls Centre for Theoretical Physics, University of Oxford, Oxford OX1 3PU, United Kingdom}

\date{\today}
\begin{abstract}
Enzymes are nano-scale machines that have evolved to drive chemical reactions out of equilibrium in the right place at the right time. Given the complexity and specificity of enzymatic function, bottom-up design of enzymes presents a daunting task that is far more challenging than making passive molecules with specific binding affinities or building nano-scale mechanically active devices. We present a thermodynamically-consistent model for the operation of such a fuelled enzyme, which uses the energy from a favourable reaction to undergo non-equilibrium conformational changes that in turn catalyze a chemical reaction on an attached substrate molecule. We show that enzymatic function can emerge through a bifurcation upon appropriate implementation of momentum conservation on the effective reaction coordinates of the low dimensional description of the enzyme, and thanks to a generically present dissipative coupling. Our results can complement the recently developed strategies for de novo enzyme design based on machine learning approaches.
\end{abstract}

\maketitle

\section{Introduction}

A fundamental question at the core of many areas of research is how to develop novel strategies for achieving non-equilibrium control at the nano-scale over the direction of chemical reactions \cite{Borsley2024}. While this is a challenging task from the point of view of bottom-up synthetic approaches, there are lessons to be learned from biological enzymes that have evolved to carry out such tasks with a high degree of precision and robustness despite the overwhelming buffeting by the environment in which they function. Since the pioneering work by Michaelis and Menten who elucidated some of the key phenomenological aspects of enzyme-assisted reaction kinetics \cite{MichaelisMenten,johnson2011original}, our understanding of the underlying mechanisms behind these processes has been progressively refined \cite{qian2007phosphorylation,Lu1998,English2005,Kou2005}. The underlying physical picture provided by the Kramers theory of noise-activated barrier crossing has provided a key conceptual framework for a reduced low-dimensional characterization of catalytic processes along appropriately selected reaction coordinates \cite{kramers1940brownian,hanggi1990reaction}, while molecular dynamics (MD) simulations of enzymes have been developing rapidly with the help of increased computing power \cite{Karplus2002}.

All-atom MD approaches give important insight at the microscopic level, for particular systems of interest. However, they deal with many degrees of freedom and simulation parameters (e.g.~atomic force fields), which might not readily lend themselves to the goal of extracting intuitive, minimal guidelines for the de novo design and optimization of enzymes. There is moreover still a large gap in time scales that can be computationally afforded to simulate sizeable numbers of catalytic cycles for most enzymes \cite{Kutzner2019}. With the advent of machine learning approaches that are increasingly used in the design and optimization of enzymes \cite{Lovelock2022,Yeh2023,Dallago2023}, it is evident that such minimal design rules, which reflect the physical constraints on the conformational dynamics of proteins while driven away from equilibrium through catalytic activity, will be of great potential promise.

In recent years, fundamental insight from statistical physics has been successfully used towards experimental demonstrations of many interesting non-equilibrium phenomena \cite{Kathan2021,Corra2022,Sorrenti2017,Pumm2022}. To aim towards the development of systems that can accomplish relatively complex tasks, one can benefit from innovative ideas that build on emergent physical properties of many-body non-equilibrium systems. Recent examples of such developments include proposals to employ non-reciprocal interactions to design collective barrier-crossing strategies that could emulate enzymatic function at larger scales \cite{Osat2024,Klapp2024}, autonomous multifarious self-organization of complex protein structures \cite{Osat2023}, as well as proposed scenarios for fast and efficient self-organization of primitive metabolic cycles at the early stages of life formation \cite{OuazanReboul2023}.

The non-equilibrium dynamics of an enzyme during catalysis simultaneously involves energy transduction and conformational changes \cite{glowacki2012taking,callender2015dynamical}, i.e. displacements. This suggests that mechanical considerations should play a key role in the stochastic dynamics of an enzyme, and consequently, in its optimal design with the aim of achieving the desired catalytic cycle. In other words, the mechanical activity of enzymes that lies at the core of their function as efficient catalysts bears a strong resemblance to the dynamics of molecular motors \cite{juelicher1997modeling,Kolomeisky2007,RevModPhys.92.025001}, with the important difference that the output energy is not in the form of mechanical work. Just like in the development of synthetic nano-motors, e.g. DNA-origami-based prototypes \cite{Pumm2022,Shi2022,Shi2023}, it would be desirable to have bottom-up strategies to build synthetic enzymes.  

Here, we set out to construct a minimal model for a `fuelled' enzyme, so that we can extract a set of golden rules regarding its optimal design (see below). Our theoretical framework is built upon two main pillars, namely, momentum conservation as a fundamental physical constraint on the effective reaction coordinates that span the relevant low dimensional configuration space of the system, and a generically present dissipative coupling between the different reaction coordinates, which we systematically derive from a microscopic model. The implementation of momentum conservation leads to the emergence of an effective mechanochemical coupling in the same spirit as models used in studies of stochastic nano-swimmers \cite{RG2010,RG2008,RG2009stoch,RG-prl2015} and enhanced diffusion of enzymes \cite{Illien2017,AgudoCanalejo2018}, while the dissipative coupling has been recently used to show how enzyme pairs can cooperate by exhibiting synchronization and enhanced catalytic activity \cite{jaime}. The two ingredients described above give rise to a bifurcation in the dynamical phase-portrait of the low dimensional configuration space, which enables enzymatic activity as an emergent feature of the dynamics, via a fundamentally novel mechanism not accessible to Kramers-like energy-barrier-crossing descriptions in terms of reaction coordinates. Typically, those descriptions are based on a modification of the energy landscape of the reaction induced by the presence of the enzyme. Here, the energy landscape remains unchanged in the presence of the enzyme, but the dynamics on this landscape becomes qualitatively different due to the cross-talk between different reaction coordinates which arises naturally as a result of the bifurcation. Our work complements other studies that involve minimal models for studying enzymatic behaviour, often based on colloids interacting through short-ranged interaction potentials or bead-spring networks \cite{zeravcic2014self,zeravcic2017spontaneous,zeravcic2017colloquium,rivoire2020geometry,McMullen2022,munoz2023computational,rivoire2023flexibility}. These studies predominantly describe the action of a `passive' enzyme, namely, catalysts that can accelerate a process of interest in the thermodynamically favourable direction. 

\begin{figure}[t]
\centering
\includegraphics[width=\linewidth]{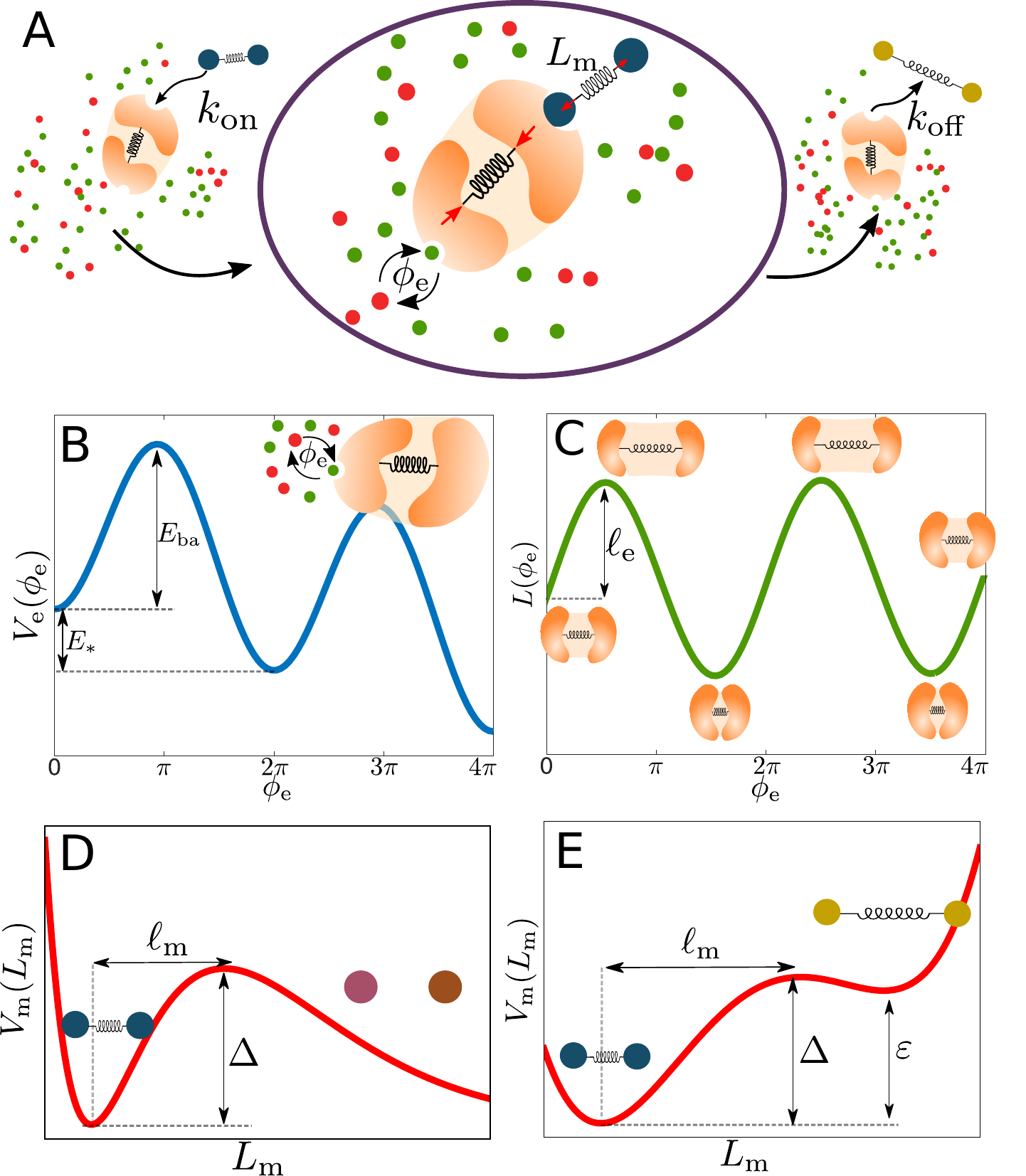}
\caption{\textbf{Minimal model of an enzymatic reaction} (A) We focus only on the catalytic step of the reaction, where a fuel-to-waste conversion ($\mathrm{\phi_\ee}$) causes a substrate-to-product reaction. For simplicity, we omit the substrate binding ($k_\mathrm{on}$) and product unbinding ($k_\mathrm{off}$) steps, assuming that they take place rapidly after each catalytic reaction. (B) The free energy landscape that drives the internal (fuel-to-waste) thermodynamically favourable reaction of the enzyme, where $E_\mathrm{ba}$ is the energy barrier height and $E_*$ the difference in the Gibbs free energy after a complete reaction. (C) The dynamics of the preferred length of the enzyme during the internal reaction, which induces conformational changes of amplitude $\ell_\ee$ during a reaction. (D, E) Free energy landscapes for the molecule, representing the thermodynamically unfavourable substrate-to-product reaction, for two examples of reaction: (D) dissociation of a dimer molecule into two monomers, (E) transition of a molecule between a short and a long conformational state. Here, $\Delta$ is the barrier height, $\varepsilon$ is the Gibbs free energy difference, and $\ell_\mm$ is the difference in the length of molecule between the initial and the transition state.  }\label{fig:figure_intro}
\end{figure}

We focus on the catalytic step of an enzymatic reaction (Figure~\ref{fig:figure_intro}(A)) and consider the conformational changes of the enzyme during the conversion of a fuel molecule to a waste molecule (Figure~\ref{fig:figure_intro}(B,C) and Supplemental Videos 1 and 2). During each reaction, the enzyme undergoes an expansion and contraction cycle. This could also serve as a simple model of an  active nanomotor that undergoes shape oscillations, as used in recent experiments \cite{Tang2025Jan}. We consider two types of substrate molecules: a dimer that can dissociate into two monomers (Figure~\ref{fig:figure_intro}(D)), and a dimer that has two states, a short and a long one (Figure~\ref{fig:figure_intro}(E) and Supplemental Video 3). In both cases, the dimer is attached at the outside of the enzyme. Through a detailed, thermodynamically consistent calculation, we show that during the fuel-induced contraction the enzyme can cause the dissociation of the dimer into monomers, or its transition from the short to the long conformation, respectively, even if these processes are not favoured thermodynamically or face a substantial energy barrier (see Supplemental Video 4). 
These catalyzed reactions are favourable in a large portion of the parameter space, related to the sizes and lengths of the enzyme and the bound molecule.

Our proposed dynamical paradigm, built on appropriate implementation of the relevant physical constraints on the minimal reaction coordinates, allows us to identify the following three golden rules for the optimal function of a fuelled enzyme driven by mechanochemical coupling: 
(i) the enzyme and the molecule should be attached at their corresponding smaller ends (i.e. higher mobility ends);
(ii) the conformational change of the enzyme must be comparable to, or larger than, the conformational change required of the molecule;
(iii) the conformational change of the enzyme must be sufficiently fast such that the molecule is stretched, rather than simply translated along with the enzyme without stretching. The rules can provide useful input to the complementary perspectives of de novo enzyme design based on machine learning and all-atom simulations.

The article is organized as follows. We first present the enzyme-molecule model. After coarse-graining, we derive the general form of the dynamical equations of the enzyme and molecule. After introducing the relevant geometric parameters and energy landscapes, we study the corresponding deterministic dynamical system. We show that the catalytic action of the enzyme can be understood as emerging, in the mathematical language of dynamical systems theory, from a novel global bifurcation in the deterministic phase-space dynamics. This bifurcation is a consequence of the non-equilibrium drive of the fuel-to-waste reaction and the mechanochemical coupling with the passive molecule arising from enzyme conformational changes. We confirm the latter through the analysis of the stochastic dynamics in the presence of thermodynamically consistent noise. By studying the non-equilibrium steady-state of the system, we show the existence of an optimal set of parameters. Finally, an analysis of first-passage times shows that fuelled catalysis enables a great reduction of the characteristic reaction time and thus a substantial enhancement in the reaction rate.

\section{Model}\label{sec2}

In biological cells, many reactions are driven by enzymes that use a thermodynamically favourable reaction (fuel-to-waste) to power a thermodynamically unfavourable reaction (substrate-to-product) that would normally (spontaneously) take place in the opposite direction, in the absence of the enzyme. The most common fuel for these reactions is ATP, which is converted into useful work or motion \cite{alberts2017molecular,RevModPhys.92.025001, Borsley2022,Astumian1996,Eide2006}. Here we develop a minimal model to describe the dynamics of a fuelled enzyme.

\subsection{Geometry and deterministic dynamics}

We start with the simple model for an enzyme that undergoes conformational changes during a fuel-to-waste reaction, as previously introduced in Ref.~\citenum{jaime}. In particular, the fuel-to-waste reaction is not modelled explicitly, and is instead described by an internal phase $\phi_\ee$ which moves stochastically along a downhill washboard potential $V_\ee(\phi_\ee)$; see Figure~\ref{fig:figure_intro}(B). A noise-activated jump over one of the energy barriers (of height $E_\mathrm{ba}$) leads to an increase of the internal phase $\phi_\ee$ by $2\pi$, and corresponds to the conversion of fuel into waste with a free energy release given by $E_*$. We assume that the fuel-to-waste process is reaction-limited, which is equivalent to fuel molecules being sufficiently abundant, such that a new fuel molecule would effectively bind instantly to the enzyme after a reaction allowing the process to be repeated again (see Figure~\ref{fig:figure_intro}(A)).

The enzyme is represented by two sub-units, with the separation $L_\ee$ between them representing the mechanical degree of freedom that is coupled to the internal chemistry. This coupling is described by the potential $U(L_\ee, \phi_\ee ) = \frac{k}{2}(L_\ee - L(\phi_\ee) )^2 + V_\ee(\phi_\ee)$,
where the first term is a harmonic potential that couples the actual length $L_\ee$ to a preferred length $L(\phi_\ee)$ that depends on the internal phase describing the fuel-to-waste reaction; see Figure~\ref{fig:figure_intro}(C). As a consequence of this coupling, every time that a fuel-to-waste reaction occurs the enzyme undergoes a cyclic conformational change.

The dynamics of the dimer molecule is described by a single degree of freedom, which is the separation $L_\mm$ between its two monomers; see Figure~\ref{fig:figure_intro}(D,E).
To model the two different chemical reactions already described (dimer dissociation and short-to-long conformation switch), two different potentials $V_\mm(L_\mm)$ that govern the length of the molecule are used. For convenience and to unify the notation in both cases, we express the dimer length as $L_\mm = L_\mm^{(0)} + \ell_\mm \phi_\mm$ where $L_\mm^{(0)}$ is the length in the dimer state or in the transition state between short and long, respectively, and $\ell_\mm$ is the length increase necessary to reach the transition state; see Figure~\ref{fig:figure_intro}(D,E).

\begin{figure*}[t]
\centering
\includegraphics[width=0.8\linewidth]{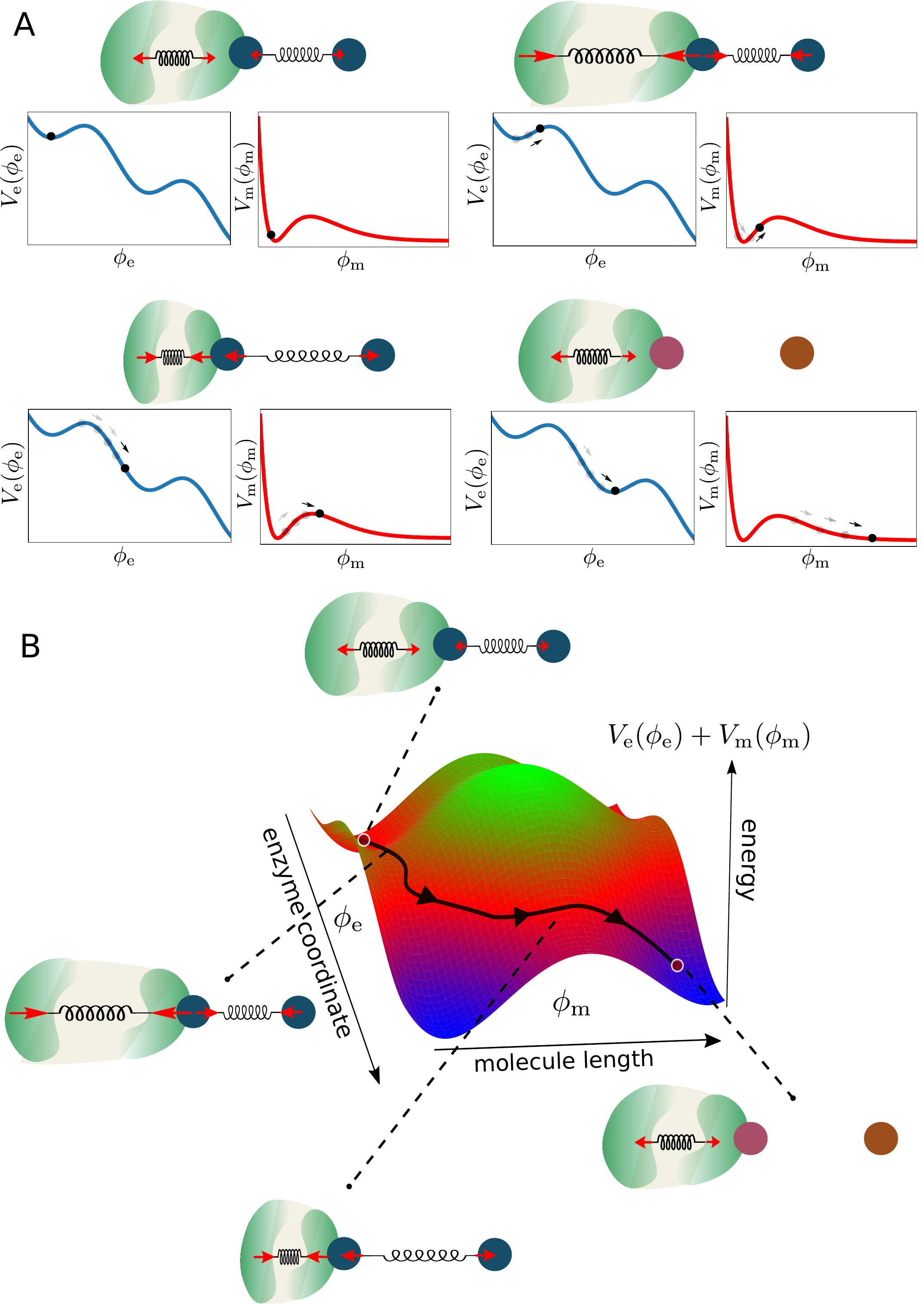}
\caption{\textbf{The evolution along the two-dimensional energy landscape} (A) A visualization of a dissociation process catalyzed by a fuelled enzyme, divided into four stages. The blue and red curves correspond to the energy landscape of the internal reaction of the enzyme $V_\ee(\phi_\ee)$, and of the molecular reaction $V_\mm(\phi_\mm)$, respectively. The solid black circles on the energy landscapes indicate the present state of each process, depicted in each respective schematic figure. The red arrows represent internal forces on the enzyme and molecule. (B) The same process visualized within the two dimensional energy landscape given by $V_\ee(\phi_\ee)+V_\mm(\phi_\mm)$, where the black line shows the evolution of the combined enzyme-molecule system. Thermal noise tends to kick the system over the enzymatic energy barrier, since it is smaller than the barrier in the energy landscape of the molecule. The coupling between the enzyme and molecule dynamics, mediated by the conformational changes of the enzyme, then causes the system to cross over the energy barrier of the molecular reaction.}\label{fig:figure2}
\end{figure*}

We now briefly describe the dynamics governing these degrees of freedom (for the detailed derivation see SI). When the substrate molecule strongly binds to the enzyme, a complex effectively made up of three sub-units is formed (Figure~\ref{fig:figure2}). Each sub-unit has a hydrodynamic mobility (inverse friction): $\mu_\ee$ for the part of the enzyme not bound to the molecule, $\mu_\bb$ for the unit that is shared by the molecule and the enzyme when bound to each other, and $\mu_\mm$ for the part of the molecule that is not bound to the enzyme. Since the typical time scales of the catalytic activity and the associated enzyme conformational transitions are much longer than the relevant inertial time scale \cite{RG-prl2015}, we consider the overdamped dynamics of these three sub-units, which leads to coupled evolution equations for the two lengths. Similarly, we consider overdamped dynamics for the evolution of the internal phase of the enzyme along the chemical free energy landscape, with an associated mobility $\mu_\phi$. By assuming that the enzyme is stiff, one can enslave the dynamics of the enzyme length (fast variable) onto the dynamics of the internal phase (slow variable). This reduces the degrees of freedom to two, and results in coupled equations for the internal phase of the enzyme and the length of the molecule. The deterministic components of the equations take on the form
\begin{equation}\label{eq:deterministic}
    \dot \phi_\alpha = \sum_{\beta=\ee,\mm} M_{\alpha \beta}(\phi_\ee) [-\partial_{\beta} V_\beta(\phi_\beta) ]
\end{equation}
where $M_{\alpha \beta}$ is a symmetric effective mobility tensor (with $\alpha,\beta \in \{\ee,\mm\}$), and $\partial_\beta \equiv \frac{\partial}{\partial\phi_\beta}$ (note that we do not implement a summation convention). The explicit expressions for the components of the mobility tensor are given as follows
\begin{align}
    M_{\ee \ee} &= \frac{\mu_\phi}{1+\frac{\mu_\phi}{\mu_1} L'(\phi_\ee)^2}, \label{eq:m11}\\
    M_{\ee \mm} &=M_{\mm \ee}= - \bigg(\frac{\mu_\phi}{1+\frac{\mu_\phi}{\mu_1} L'(\phi_\ee)^2}\bigg)\frac{\mu_\bb}{\mu_1}\frac{L'(\phi_\ee)}{\ell_\mm},\label{eq:off-diag}\\
    M_{\mm \mm}&=\frac{1}{{\ell_\mm}^2}\bigg(\mu_2 - \frac{\mu_\bb^2}{\mu_1}\frac{1}{1+\frac{\mu_\phi}{\mu_1} L'(\phi_\ee)^2}\bigg),\label{eq:m22}
\end{align}
where $\mu_1 \equiv \mu_\ee + \mu_\bb$, $\mu_2\equiv\mu_\mm + \mu_\bb$, and $L'(\phi_\ee) = \frac{d L(\phi_\ee)}{d\phi_\ee}$. 

\subsection{The effect of noise in the stochastic dynamics}

Since enzymes operate at the nano-scale, they are strongly affected by the thermal fluctuations of the surrounding medium. In particular, because the dynamics is dissipative, thermal noise is essential to kick the system out of local energy minima and over barriers associated with the reactions; see Figure~\ref{fig:figure_intro}(B,D,E).

Therefore, it is essential to introduce thermal fluctuations in our description of the problem. 
The system is out of equilibrium due to the driving force given by the energy $E_*$ of the fuel. In the limit of $E_*=0$, the steady state of the system must be a state of thermodynamic equilibrium, corresponding to a Boltzmann distribution. This implies that the stochastic dynamics corresponding to the deterministic dynamics in Equation~\eqref{eq:deterministic} must be described by a Fokker-Planck equation describing the time evolution of the probability distribution $P(\phi_\ee,\phi_\mm,t)$ as follows
\begin{align}
    \partial_t P + \sum_{\alpha=\ee,\mm} \partial_\alpha  J_\alpha= 0, \label{eq:fokker-planck} 
\end{align}
with the conserved probability currents given as
\begin{align}
    J_\alpha (\phi_\ee,\phi_\mm,t) =\sum_{\beta=\ee,\mm} M_{\alpha \beta}\big[P (-\partial_\beta V_\beta) - k_{\rm B} T \partial_\beta P \big], \label{eq:currents}
\end{align}
where $k_{\rm B}$ is the Boltzmann constant and $T$ is the temperature. In the absence of a non-equilibrium driving force ($E_*=0)$, both probability currents $J_\mathrm{e}$ and $J_\mathrm{m}$ must vanish at the equilibrium steady state. As a result, one recovers the Boltzmann distribution $P_\mathrm{eq} \propto \exp(-V_\ee(\phi_\ee)/k_{\rm B} T)\exp(-V_\mm(\phi_\mm)/k_{\rm B} T)$ independently of the form of the mobility matrix $M_{\alpha \beta}$, and thus of the geometric parameters describing the enzyme and the molecule.

That the equilibrium probability distribution factorizes reflects the fact that the two-dimensional energy landscape $V_\ee(\phi_\ee) + V_\mm(\phi_\mm)$ in which the phases evolve is separable, i.e.~there is no coupling between $\phi_\ee$ and $\phi_\mm$ through the potential. The dynamics of $\phi_\ee$ and $\phi_\mm$ are coupled only through the off-diagonal components of the mobility tensor $M_{\alpha \beta}$, which arise due to the mechanical contact (binding) between enzyme and molecule in combination with the conformational changes of the enzyme, and represent a form of \textit{dissipative} coupling which only plays a role in a non-equilibrium setting ($E_*>0$), i.e.~when the fuel-to-waste reaction is thermodynamically favoured. This type of coupling has been seen to emerge in models of coupled molecular machines (\emph{via} mechanical binding or hydrodynamic interactions mediated by a viscous fluid) and enables cooperative phenomena like synchronization \cite{jaime,chatzittofi} and phase-locking \cite{chatzittofi2023topological} among enzymes, of which there is suggestive experimental evidence \cite{Lu2014,Hehir2000}.

The Langevin equation associated to the Fokker-Planck equation in Equation~\eqref{eq:fokker-planck} takes the following form in the Stratonovich convention,
\begin{align}
    \dot \phi_\alpha = \sum_{\beta=\ee,\mm} & \Big[M_{\alpha \beta} (-\partial_\beta V_\beta )  + \sqrt{2 k_{\rm B} T} \sigma_{\alpha \beta} \xi_{\beta} \nonumber \\
    &+  \sum_{\nu=\ee,\mm} k_{\rm B} T \sigma_{\alpha \nu} \partial_\beta \sigma_{\beta \nu} \Big],\label{eq:langevin}
\end{align}
where $\sum_{\nu=\ee,\mm} \sigma_{\alpha \nu} \sigma_{\beta \nu} = M_{\alpha \beta}$. The term on the second line of the right hand side is the spurious drift term arising from the multiplicative nature of the noise, due to the dependence of the mobility matrix on $\phi_\ee$, while the last term on the first line of Equation \eqref{eq:langevin} is the thermal noise with $\xi_\beta$ being a unit white noise such that $\langle \xi_\beta(t)\rangle = 0$ and $\langle \xi_\alpha(t) \xi_\beta(t') \rangle = \delta_{\alpha \beta} \delta(t-t')$. It is also worth noting that the mobility matrix must be symmetric and positive definite (which is the case here) in order for the dynamics to be thermodynamically consistent \cite{degroot,kim2013microhydrodynamics}.

\begin{table*}[t]
  \begin{center}
    \begin{tabular}{cccl} 
    \hline\hline
     Parameters & \;\;\;\;\; & \;\;\;\;\; & Description\\
      \hline\hline
    $\mu_\phi$ & & & mobility of the internal phase $\phi_\ee$ (reaction coordinate of fuel-to-waste reaction)\\
    $\mu_\ee$ & & & hydrodynamic mobility of the enzyme-only sub-unit \\
    $\mu_\mm$ & & & hydrodynamic mobility of the molecule-only sub-unit \\
    $\mu_\bb$ & & & hydrodynamic mobility of the shared (bound) sub-unit \\
    $\mu_1 = \mu_\ee + \mu_\bb$ & & & sum of the mobilities of the enzyme-only and shared sub-units \\
    $\mu_2=\mu_\mm + \mu_\bb$ & & & sum of the mobilities of the molecule-only and shared sub-units \\
    $h=\mu_\bb/(\mu_\ee + \mu_\bb)$ & & & effective coupling strength ($0 \leq h \leq 1$) \\
    $\ell_\ee$ & & & amplitude of the conformational change of the enzyme \\
    $\ell_\mm$ & & & length difference between the initial and transition states of the molecule\\
    $E_\mathrm{ba}$ & & & energy barrier of the fuel-to-waste reaction\\
    $E_*$ & & & Gibbs free energy released by the fuel-to-waste reaction\\
    $\Delta$ & & & energy barrier of the molecule reaction\\
    $\varepsilon$ & & & unfavourable energetic bias of the short-to-long reaction\\
    $k_{\rm B} T$ & & & thermal energy (strength of thermal fluctuations)\\
      \hline \hline
    \end{tabular}
  \end{center}
  \caption{Summary of the parameters of the model.\label{tab:table1}}
\end{table*}

\subsection{Model parametrization}

The enzyme and the molecule are embedded in a solvent under low Reynolds number conditions \cite{kim2013microhydrodynamics}. Therefore, we can infer that the mobilities $\mu_\ee$ (enzyme-only sub-unit), $\mu_\bb$ (bound enzyme-molecule sub-unit), and $\mu_\mm$ (molecule-only sub-unit) are related to their corresponding effective hydrodynamic radii $a_\ee$, $a_\bb$, and $a_\mm$, via the Stokes relation $\mu=1/(6\pi \eta a)$, ignoring hydrodynamic interactions in this lowest order approximation. The mobilities thus directly encode information about the sizes of the enzyme and the molecule sub-units. The length scale related to the amplitude of the deformation of the molecule is $\ell_\mm$; see Figure~\ref{fig:figure_intro}(D,E). For the conformational changes of the enzyme, we assume that the preferred length $L(\phi_\ee)$ oscillates during the enzymatic cycle as 
\begin{align}
    L(\phi_\ee) = L_\ee^{(0)} + \ell_\ee \sin(\phi_\ee - \delta),
    \label{eq:Lphi-ee}
\end{align}
where $L_\ee^{(0)}$ is an average length which does not enter the dynamics, $\ell_\ee$ is the amplitude of the conformational change, and $\delta$ is a phase shift described below (we choose $\ell_\ee\geq 0$).

The washboard potential governing the dynamics of the internal phase $\phi_\ee$ is taken to be of the form $V_\ee(\phi_\ee) = -F \phi_\ee + v \sin(\phi_\ee - \delta)$ where $\delta = \arccos(F/v)$ is the value of the phase shift which guarantees that the minima of the potential are located at integer multiples of $2 \pi$. The parameters $F$ and $v$ can be mapped to the height of the energy barrier of the enzymatic reaction $E_{\mathrm{ba}}$ and the free energy released after a complete reaction $E_*$ (see Figure~\ref{fig:figure_intro}(B)) through $E_{\mathrm{ba}} = [2 \sqrt{1-(F/v)^2}+2 \delta(F/v)] v$ and $E_*= 2 \pi F$ \cite{jaime,chatzittofi2023topological,Chatzittofi2024Aug}. By eliminating $v$ and $F$, we find that the parameter $\delta$ is given by the solution of the transcendental equation 
\begin{align}
\delta + \tan\delta = \pi E_\mathrm{ba}/E_*, \label{eq:transcend}
\end{align}
which yields the approximate solution
\begin{align}
    \delta \simeq   \frac{\pi}{2} \frac{E_\mathrm{ba}}{E_*},\label{eq:delta-apprx}
\end{align}
for small values of $E_\mathrm{ba}/E_*$. 

The parameter $\mu_\phi$ is the mobility of the internal phase $\phi_\ee$ and may have different microscopic origins depending on the particulars of the fuel-to-waste reaction. In principle, it can be related to the rate $k(\mathrm{f} \to \mathrm{w})$ of the fuel-to-waste reaction and the height of the energy barrier $E_{\mathrm{ba}}$ through the Kramers escape rate \cite{kramers1940brownian,hanggi1990reaction} 
\begin{align}
    k(\mathrm{f} \to \mathrm{w}) = \frac{\mu_\phi}{2 \pi}\sqrt{\lvert\lambda_\mathrm{e}^{(1)}\rvert \lambda_\mathrm{e}^{(0)}} \exp\bigg(-\frac{E_\mathrm{ba}}{k_{\rm B} T}\bigg), \label{eq:kramers}
\end{align}
where $\lambda_\mathrm{e}^{(0)}=V_\ee''(\phi_\mathrm{e})|_{\rm min}$ and $\lambda_\mathrm{e}^{(1)}=V_\ee''(\phi_\mathrm{e})|_{\rm max}$, i.e. they correspond to the values of the second derivative (curvature) of the enzyme potential around the minimum and the maximum (energy barrier), respectively. Thus, by knowing the reaction rate and barrier height one can infer the value of $\mu_\phi$.

For the molecule, we consider two different types of reaction starting from a dimer-like substrate. For the dissociation reaction, we consider a potential with a minimum representing the stable initial state of the molecule, a maximum corresponding to an energy barrier with height $\Delta$, and an asymptotic decrease to a constant value, which represents the fact that beyond the barrier the monomers become disconnected.
We describe this with the potential $V_\mm(\phi_\mm) = \Delta \phi_\mm^2 \exp[2(1-\phi_\mm)]$, shown in Figure~\ref{fig:figure_intro}(D).
For the conformational switch reaction where the dimer switches from a short to a long state, we choose a bistable potential, with an energy barrier $\Delta$ and an energy difference $\varepsilon$ between the long (higher energy) and the short (lower energy) states. We model this using the potential $V_\mm(\phi_\mm) = \Delta\phi_\mm^2 (\phi_\mm^2 - 2) +\varepsilon \phi_\mm/2$, with $\varepsilon <\Delta $; see Figure~\ref{fig:figure_intro}(E).
All the relevant model parameters are summarized in Table~\ref{tab:table1}.

\begin{figure*}[t]
\centering
\includegraphics[width=0.85\linewidth]{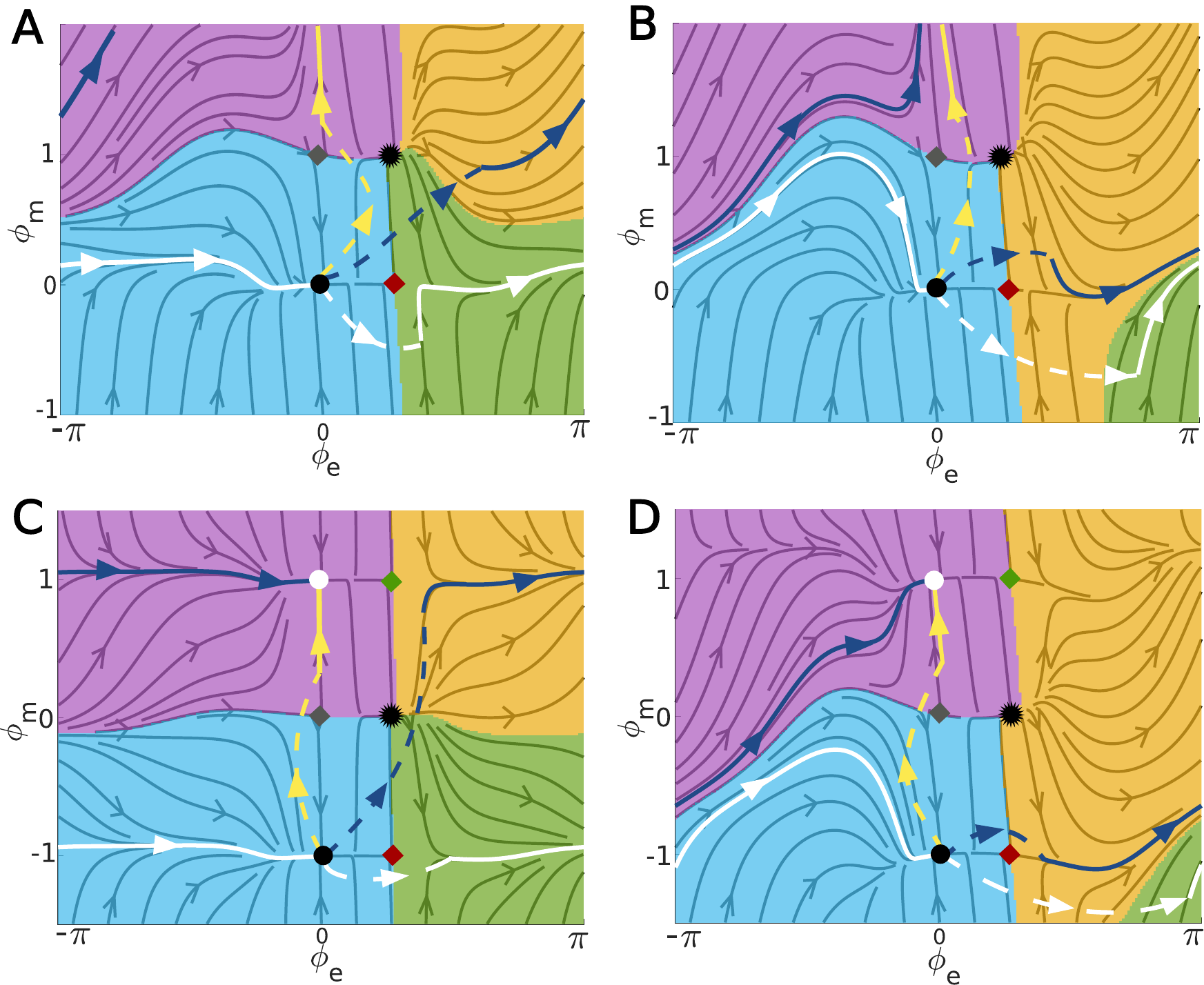}
\caption{\textbf{Phase-portraits of the deterministic dynamics}  (A,B) the dissociation reaction and (C,D) the switch reaction. A global bifurcation occurs with increasing coupling strength $h$, which is below the critical value in (A,C) and above it in (B,D). In all cases, stable fixed points are depicted as circles, saddle points as diamonds, and unstable fixed points as asterisks. Dashed lines depict possible scenarios in which thermal fluctuations can kick the system out of the basin of attraction of a stable fixed point. The solid flow lines show typical trajectories, and their final resting points, when starting from different basins of attraction. 
The different colouring of the diamonds is described in the text. The values of the coupling are (A)  $h=0.33$, (B) $h=0.48$, (C) $h=0.12$, and (D) $h=0.48$. }\label{fig:deterministic_dynamics}
\end{figure*}

\section{Results}

\subsection{Preliminary considerations}

It is crucial to note that the components of the mobility tensor $M_{\alpha \beta}$ are determined by the  geometric features of the enzyme and molecule, which can be tuned as to optimize the function of the enzyme. The basic principle is illustrated in Figure~\ref{fig:figure2}(A), for the example of dissociation of a dimer molecule. When a fuel-to-waste reaction takes place (the internal phase $\phi_\ee$ is kicked over the chemical energy barrier by thermal noise), the coupling to the molecule length arising from the conformational changes of the enzyme drives a corresponding evolution of the molecule length, as represented by $\phi_\mm$, and in particular can cause $\phi_\mm$ to overcome its energy barrier. The evolution along the two-dimensional energy landscape $V_\ee(\phi_\ee)+V_\mm(\phi_\mm)$ is shown in Figure~\ref{fig:figure2}(B), where the black path corresponds to a catalyzed reaction.

We expect to have a strong coupling between the fuel-to-waste ($\phi_\ee$) and substrate-to-product ($\phi_\mm$) reactions when the off-diagonal components in the mobility tensor are non-negligible.
In the limiting cases where either the middle sub-unit is large compared to the enzyme-only sub-unit, $\mu_\bb/\mu_1 \rightarrow 0$, or the conformational changes of the enzyme are small compared to the dissociation length of the molecule, $\ell_\ee/\ell_\mm \rightarrow 0$, the off-diagonal terms vanish, leading to a decoupling of the dynamics between the enzyme and the molecule. To achieve a strong coupling, we would aim to maximize $\mu_\bb/\mu_1$ and $\ell_\ee/\ell_\mm$, from which one can infer the first two golden rules [(i) and (ii)] for designing an enzyme discussed above. In the following, we define the ratio $h\equiv\mu_\bb/\mu_1=\mu_\bb/(\mu_\ee+\mu_\bb)$ to be the ``coupling strength'', and we note that it is bounded as $0 \leq h \leq 1$. For $h>0.5$, the mobility of the shared sub-unit is larger than that of the enzyme-only sub-unit, and the reverse is true for $h<0.5$. To achieve a fast conformational change and thus, infer the rule (iii), we would aim to maximize the chemical driving force of the fuel-to-waste process $E_*$ or increase the mobility $\mu_\phi$ governing this process.

Throughout the rest of the text we will use the following parameters unless mentioned otherwise: $\mu_\phi \ell^2_\mm/\mu_2 =1$, $\ell_\ee/\ell_\mm=2.5$, $\mu_1/\mu_2=0.88$, $E_*/\Delta = 14\pi$, $E_\mathrm{ba}/\Delta = 0.4$, and (for the switch reaction) $\varepsilon/\Delta =0.1$. The dimensionless time is defined as $\tilde t = (\mu_\phi \Delta) t$. The remaining dimensionless parameters that we explore in the text are the coupling strength $h$ and the dimensionless thermal noise strength $k_{\rm B} T/\Delta$.

\subsection{Deterministic dynamics and global bifurcation}

Before considering the full stochastic (noise-activated) dynamics of the system, we first study the deterministic dynamics given by Equation~\eqref{eq:deterministic}, as we can apply tools from dynamical systems theory to obtain an intuitive understanding of the expected behaviour, particularly in the experimentally-relevant case of low noise, where $k_{\rm B} T$ is smaller than all other energy scales.

In Figure~\ref{fig:deterministic_dynamics}, phase portraits of the deterministic dynamics are shown for low and high coupling strengths and both types of reaction (see Methods for details of the numerical calculation of phase portraits). Depending on the type of reaction, different sets of fixed points appear in the dynamics (stable fixed points are depicted as circles, saddle points as diamonds, and unstable fixed points as asterisks).
For the dissociation reaction, there are four fixed points (one stable, two saddle points, one unstable) since there is no local minimum of the molecule potential after the energy barrier. The stable fixed point at $\phi_\mm=0$ corresponds to the substrate (dimer) state, whereas $\phi_\mm \to \infty$ corresponds to the product (dissociated) state.
For the switch reaction, there are six fixed points (two stable, three saddle points, one unstable). The stable fixed point at $\phi_\mm=-1$ corresponds to the substrate (short) state, whereas that at $\phi_\mm = +1$ corresponds to the product (long) state. 
It is worth noting that the positions of the fixed points do not depend on the coupling, but only on the potentials $V_\ee(\phi_\ee)$ and $V_\mm(\phi_\mm)$. Thus, changes in the geometry of the enzyme-molecule complex (and thus in the coupling strength) leave the location and nature of the fixed points unchanged. As a consequence, there cannot be any local bifurcations in the dynamical system and only global bifurcations are allowed \cite{jaime}. By considering the topology of the different basins of attraction in the dynamical system, we note that global bifurcations occur between Figures \ref{fig:deterministic_dynamics}(A) and \ref{fig:deterministic_dynamics}(B) for the dissociation reaction, and between Figures \ref{fig:deterministic_dynamics}(C) and \ref{fig:deterministic_dynamics}(D) for the switch reaction, respectively. 

The phase portraits allow us to uncover the different possibilities emerging in the full stochastic dynamics by classifying the different types of possible deterministic trajectories, depicted as solid flow lines in Figure~\ref{fig:deterministic_dynamics}, and considering the likelihood of their stochastic activation by random thermal kicks, depicted as dashed flow lines in Figure~\ref{fig:deterministic_dynamics}, which will take the system from an initial state around a stable fixed point to the domain of interest. Trajectories starting in the cyan region in Figure~\ref{fig:deterministic_dynamics} are attracted to the fixed point at $(\phi_\ee,\phi_\mm)=(0,0)$ (for the dissociation reaction) or $(0,-1)$ (for the switch reaction). On the other hand, those starting in the green region are attracted to the fixed point at $(\phi_\ee,\phi_\mm)=(2\pi,0)$ (dissociation) or $(2\pi,-1)$ (switch). This class of trajectories, which are depicted as solid white flow lines in Figure~\ref{fig:deterministic_dynamics}, represent the process in which the molecule remains in the substrate state and the enzyme performs a futile cycle. Analogously, trajectories starting in the purple region end at $(0,\infty)$ (dissociation) or $(0,1)$ (switch), corresponding to spontaneous transition of the molecule to the product state, without an accompanying transition in the enzyme, while those in the gold region converge to $(2\pi,\infty)$ (dissociation) or $(2\pi,1)$ (switch), representing a process in which the molecular transition coincides with the enzymatic cycle. The relative frequency of the activation of the different processes corresponding to these distinct classes will depend on the likelihood of the availability of the thermal kick of the right strength that is needed to initiate each of them.

\begin{figure}[t]
\centering
\includegraphics[width=0.99\linewidth]{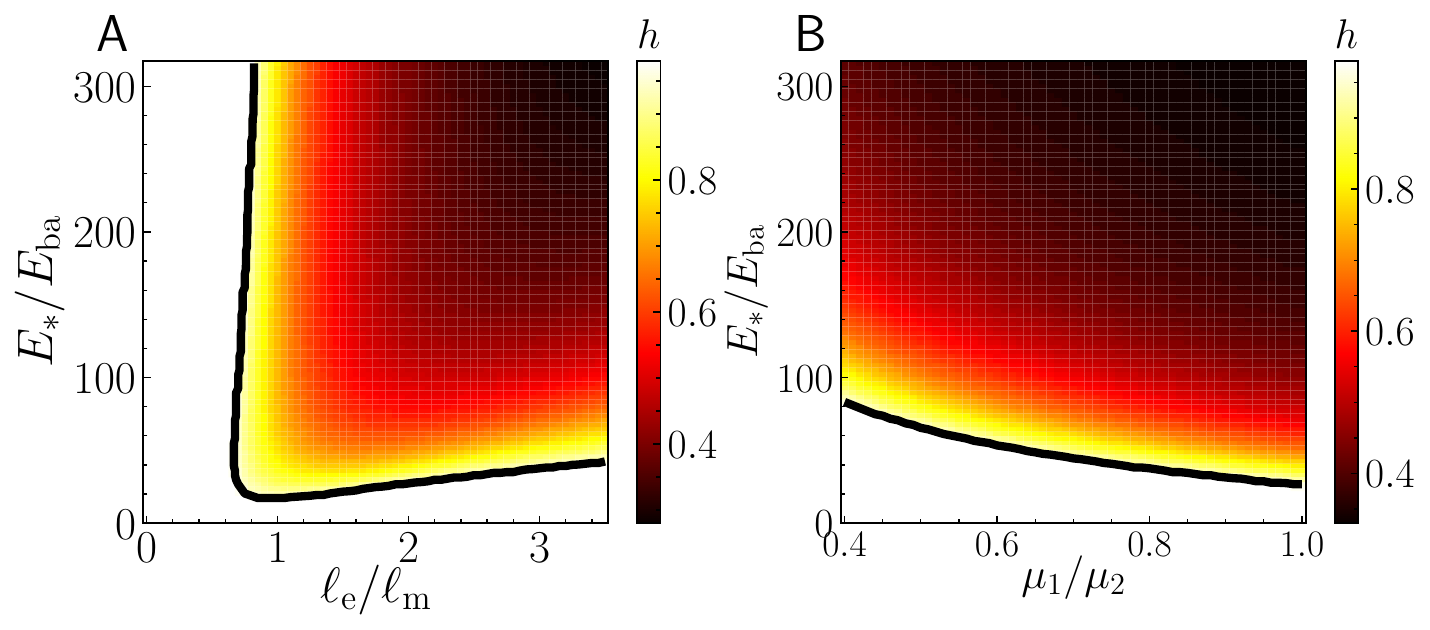}
\caption{\textbf{Critical coupling strength at which the bifurcation occurs} (A) as a function of the geometric ratio $\ell_\ee/\ell_\mm$ and the ratio $E_*/E_\mathrm{ba}$ describing the internal enzyme reaction, and (B) the mobility ratio $\mu_1/\mu_2$ and $E_*/E_\mathrm{ba}$. In the white region outside the black solid line, a bifurcation does not occur within the physical range $0 \leq h \leq 1$.}\label{fig:parspace_diagrams}
\end{figure}

\begin{figure*}[t]
\centering
\includegraphics[width=0.9\linewidth]{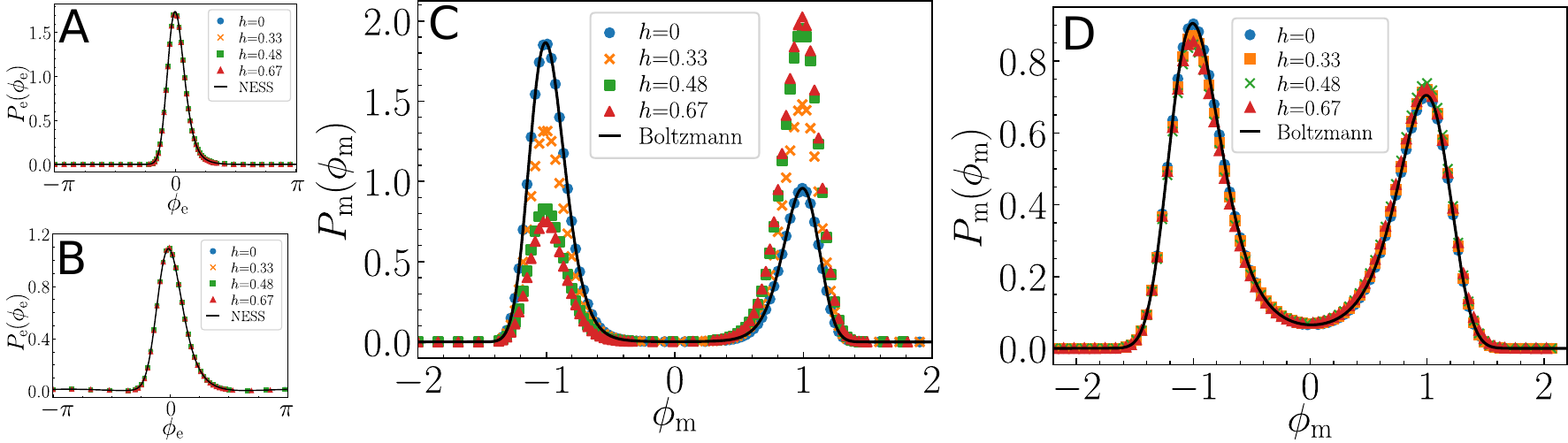}
\caption{\textbf{Marginal steady-state distributions for the switching reaction} (A,C) at low noise $k_{\rm B} T /\Delta = 0.15$, and (B,D) at high noise $k_{\rm B} T / \Delta = 0.4$. The distribution for the enzyme coordinate $\phi_\ee$ is shown in (A,B), and that for the molecule coordinate $\phi_\mm$ in (C,D).  In (A,B) the black line labelled ``NESS" is the analytical result for an uncoupled enzyme given in Equation~(S9). In (C,D) the black line labelled ``Boltzmann" is the equilibrium distribution $P_\mm(\phi_\mm) \propto \exp(-V_\mm/k_{\rm B} T)$. }\label{fig:ss_distributions}
\end{figure*}

More specifically, let us consider the case of weak coupling in Figure~\ref{fig:deterministic_dynamics}(A,C). Thermal fluctuations can most likely kick the system over one of the two saddle points or transition states, either the one to the right of the stable fixed point shown as red diamond (white trajectory)
or the one above it
shown as grey diamond (yellow trajectory).  
When traversing the red diamond, the enzyme will complete an internal reaction ending at $\phi_\ee=2\pi$, while the molecule remains in the substrate state. When traversing the grey diamond, the molecule undergoes a reaction and turns into product, while the enzyme completes no internal reaction and remains at $\phi_\ee=0$. Thus, the internal (fuel-to-waste) reaction of the enzyme and the molecule (substrate-to-product) reaction are completely uncoupled in these cases. Less likely thermal fluctuations can kick the system across the unstable fixed point shown as an asterisk (blue trajectory), leading to a simultaneous occurrence of both the enzyme and the molecule cycles. In this weak coupling regime, traversing either of the saddle points (red or grey diamonds) leads to effectively one-dimensional dynamics [see Figure~\ref{fig:deterministic_dynamics}(A,
C)] since only one of the two coordinates advances. Thus, dynamics in this regime are analogous to a Kramers-type one-dimensional barrier-crossing process.

We next consider the case of strong coupling, after the global bifurcation has occurred and the basins of attraction have rearranged; see Figure~\ref{fig:deterministic_dynamics}(B,D) and Supplemental Videos 5 and 6. Thermal fluctuations kicking the system over the grey diamond (yellow trajectory) will still result in molecular reactions without associated internal reactions in the enzyme. However, if the system is activated over the red diamond, the more likely scenario will be for the enzyme to complete an internal reaction simultaneously with the molecular reaction (blue trajectory) unlike in the weak coupling case. The two reactions thus become coupled, and in particular, a fuel-to-waste reaction inside the enzyme now triggers a substrate-to-product reaction for the molecule. The futile enzyme cycle (white trajectory) is the more unlikely scenario in this case. Importantly, the dynamics become two-dimensional since the two reaction coordinates become strongly coupled and influence each other. This occurs even if the underlying energy landscape remains unchanged, through the emergence of cross-mobility terms in the dynamics of the reaction coordinates. Thus, the present dynamical systems theory approach is crucial in understanding catalysis in this system. We note that a similar bifurcation occurs for the reverse switching process, where the energetically favourable and initial state for the molecule is the long conformation; see Supplemental Information and Figure S1. 

The bifurcation occurs above a critical value of the coupling $h$, which can be tuned by the geometric properties of the enzyme-substrate complex. Moreover, in order to have an enzyme-triggered reaction, transitions through the red diamond should be favoured over transitions through the grey diamond. This is expected to happen when the energy barrier for the internal reaction is smaller than that of the molecular reaction ($E_\mathrm{ba} < \Delta)$. It is also worth mentioning that, even though the dissociation and switch reactions involve rather different energy landscapes, the bifurcation takes place at the same set of parameters since in both cases the barrier for the molecular reaction is fixed to $\Delta$.

The occurrence of the bifurcation is crucial for the coupling between the two processes to emerge, and thus for the enzyme to become active. We thus explore how the critical coupling depends on the geometric and energetic parameters of the enzyme-substrate complex.
In Fig~\ref{fig:parspace_diagrams}, we show the critical value of the coupling $h$ as a function of the ratio $E_*/E_\mathrm{ba}$ of energies in the fuel-to-waste reaction, and $\ell_\ee/\ell_\mm$ or $\mu_1/\mu_2$ which are directly related to the geometrical properties of the complex. 
In Fig~\ref{fig:parspace_diagrams}, it is observed that the bifurcation takes place at low coupling strengths when the driving $E_*$ is large, which is linked to higher reaction coordinate speed $\dot \phi_\ee$ and therefore a faster conformational change of the enzyme, in connection with the golden rule (iii). In addition, there is a non-trivial behaviour as a function of the length ratio $\ell_\ee/\ell_\mm$ at a fixed value of $E_*/E_\mathrm{ba}$, showing non-monotonic behaviour with a minimum at intermediate values.  In Fig~\ref{fig:parspace_diagrams}(B), we see that the bifurcation is favoured at higher values of the ratio $\mu_1/\mu_2$. 
The results in Figure~\ref{fig:parspace_diagrams} serve to highlight the different parameters that must be tuned in order to favour a bifurcation in the deterministic dynamics. Below, we show how these results translate to the full noise-activated stochastic dynamics.

\begin{figure*}[t]
\centering
\includegraphics[width=0.8 \linewidth]{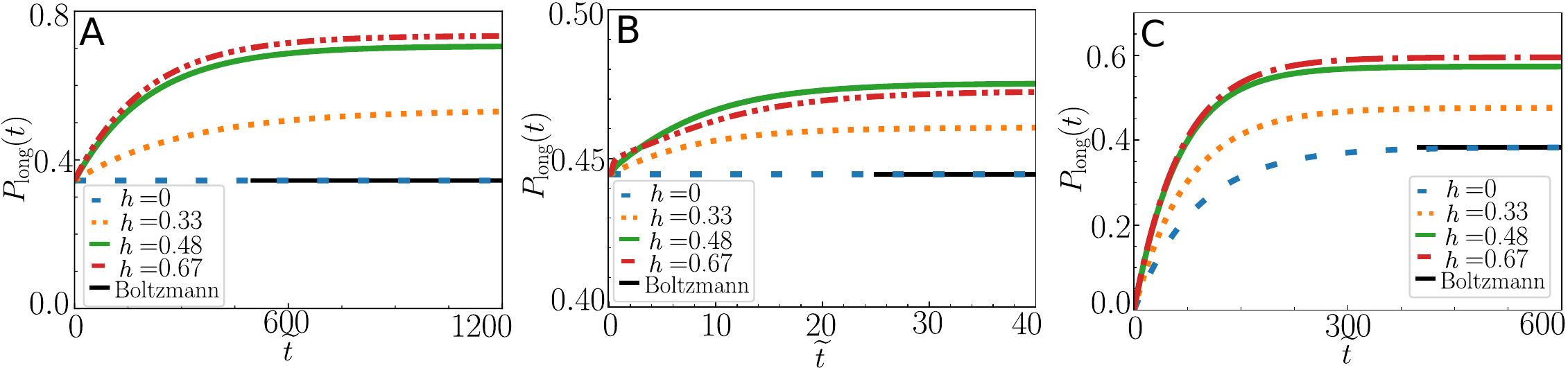}
\caption{\textbf{Probability of finding the shape-switching molecule in the long state as a function of time, for various values of the coupling $h$.} In (A,B) the initial state is chosen as a Boltzmann distribution, while in (C) it is chosen as a narrow Gaussian centered around the short state. The strength of the noise is  (A) $k_{\rm B} T /\Delta = 0.15$, (B) $k_{\rm B} T /\Delta = 0.4$, and  (C) $k_{\rm B}T/\Delta=0.2$. The horizontal line labelled ``Boltzmann" is the long-time limit expected from an equilibrium distribution.}\label{fig:plong_plots}
\end{figure*}

\subsection{Stochastic dynamics}\label{sec4}

\subsubsection{Steady state: driving a reaction uphill}

Passive enzymes can lower reaction barriers, but do not shift reaction equilibria, which are still purely governed by the free energy differences between substrate and product. Fuelled enzymes, on the other hand, can use the free energy provided by the fuel to drive the substrate-to-product reaction uphill, i.e.~favouring the formation of a high free energy product from a low free energy substrate.
To study whether our model enzyme is capable of driving reactions uphill, we focus on the switch reaction (Figure~\ref{fig:figure_intro}(E)), and numerically solve the Fokker-Planck equation (Equation~\eqref{eq:fokker-planck}, see Methods for details of the numerical solution), where we are interested in how the probability distribution evolves over time and in the form of the steady state distribution in the long time limit, where $\partial_t P_{\mathrm{ss}} \rightarrow 0 $.

In Figure~\ref{fig:ss_distributions}, the steady state marginal distributions $ P_{\ee}(\phi_\ee) \equiv \int^\infty_{-\infty} d\phi_\mm P_\mathrm{ss}(\phi_\ee,\phi_\mm)$ and $ P_{\mm}(\phi_\mm) \equiv \int^{2\pi}_0 d\phi_\ee P_\mathrm{ss}(\phi_\ee,\phi_\mm)$ for the enzyme and molecule reaction coordinates are presented for low and high thermal noise.
Both at low and at high noise, the effect of the coupling $h$ is negligible with respect to the non-equilibrium steady state (NESS) attained by the enzyme, $P_{\ee}(\phi_\ee)$; see Figure~\ref{fig:ss_distributions}(A,B). The form of $P_{\ee}(\phi_\ee)$ can be obtained analytically in the $h=0$ case (solid black line), by extending a classical calculation \cite{risken1996fokker,PhysRevLett.87.010602,PhysRevE.65.031104} to the case with multiplicative noise, i.e.~a phase-dependent mobility, as described in Supplemental Information. At high temperatures, the multiplicative noise causes a small maximum at intermediate values of $\phi_\ee$ that would not be present for additive noise; see Figure~\ref{fig:ss_distributions}(B).

In stark contrast, at low noise the coupling $h$ has a substantial effect on the NESS attained by the molecule, $P_{\mm}(\phi_\mm)$; see Figure~\ref{fig:ss_distributions}(C). At $h=0$, the distribution $P_{\mm}(\phi_\mm)$ corresponds to a Boltzmann distribution $\propto e^{-V_\mm(\phi_\mm)/k_{\rm B} T}$, and thus shows two peaks (corresponding to the short and long states), with the peak corresponding to the lower energy short state being much higher. As the coupling $h$ increases, we see this trend progressively reverse, as the fuelled action of the enzyme drives more and more probability towards the long state. This demonstrates that the fuelled action of the enzyme can indeed drive a reaction uphill, in the thermodynamically unfavourable direction. However, the action of the enzyme becomes much less effective at high noise, where deviations from the Boltzmann equilibrium are minimal; see Figure~\ref{fig:ss_distributions}(D).

\begin{figure}[t]
\centering
\includegraphics[width=0.99 \linewidth]{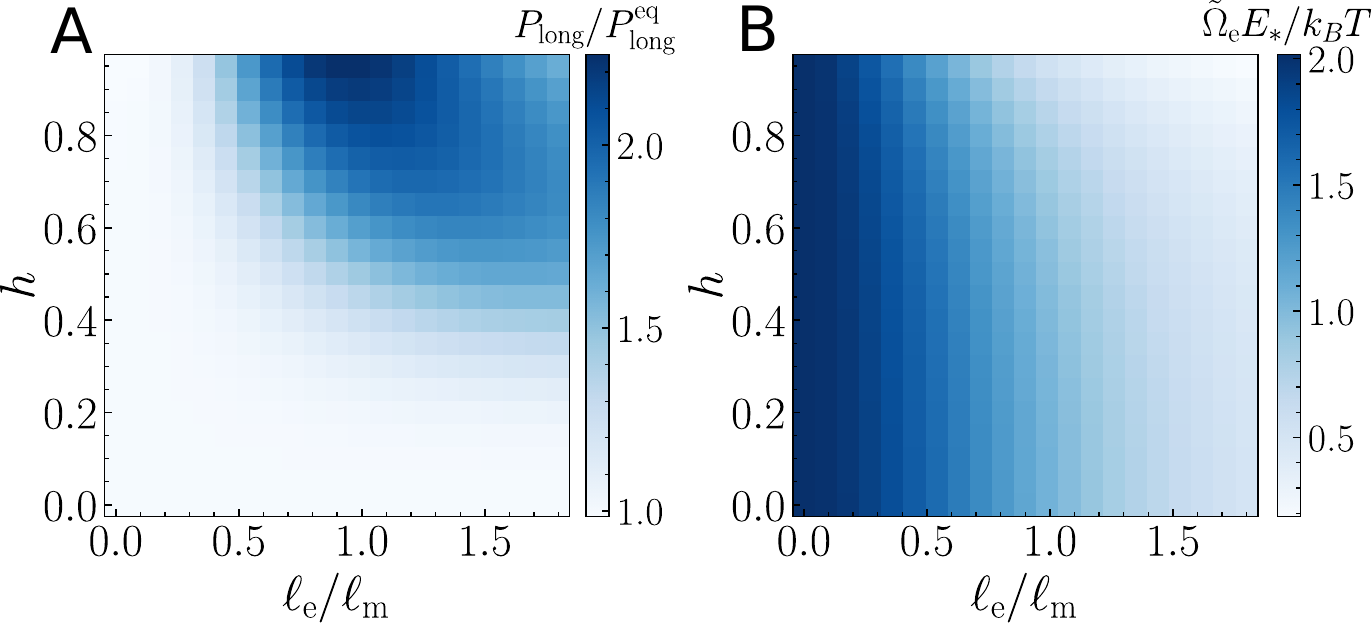}
\caption{\textbf{Results on steady-state distributions and energy dissipation} (A) The normalized probability of finding the shape-switching molecule in the long state, where $P_\mathrm{long}^\mathrm{eq}\simeq 0.383$ is the probability at equilibrium of finding the shape-switching molecule in the product state. (B) Energy dissipation rate, versus the length ratio $\ell_\ee/\ell_\mm$ and the coupling strength $h$. The noise strength is $k_{\rm B} T/\Delta = 0.2$.  }\label{fig:efficiency_ss}
\end{figure}

To further understand this effect, we consider the  time evolution of the probability of finding  the molecule in the long state $P_\mathrm{long}(t)$, defined as
\begin{align}\label{eq:plong}
    P_{\mathrm{long}}(t) = \int^{2 \pi}_0 d \phi_\ee \int^\infty_{\phi_{\mathrm{m},\mathrm{max}}} d \phi_\mm P(\phi_\ee, \phi_\mm, t),
\end{align}
where $\phi_{\mathrm{m},\mathrm{max}}$ is the location of the energy barrier of the molecular reaction along the $\phi_\mm$ coordinate.
We consider an enzyme and a molecule that are initially separated from each other, and bind at time $t=0$.  We thus initialize the enzyme probability distribution as being in the steady state of an uncoupled enzyme ($h=0$). For the  molecule, we consider two cases: (i) the molecule starts in its Boltzmann equilibrium, or (ii) the molecule starts only in the short state (with probability distributed in a narrow Gaussian around it).
Results from numerical solution of the Fokker-Planck equation (Equation~\eqref{eq:fokker-planck}) are shown in Figure~\ref{fig:plong_plots}.
For $h=0$, $P_\mathrm{long}$ tends to the probability corresponding to the equilibrium Boltzmann distribution, as expected. However, for $h>0$,   $P_\mathrm{long}$ is consistently higher than what the Boltzmann distribution dictates. Clearly, coupling to the fuelled enzyme can drive the molecular reaction uphill. The effect is most pronounced at larger coupling strengths and lower noise.
Interestingly,  $P_\mathrm{long}$ can change nonmonotonically with $h$, as seen in Figure~\ref{fig:plong_plots}(B), where $h=0.67$ shows a smaller $P_\mathrm{long}$ than $h=0.48$. This is most clearly seen in Figure~\ref{fig:efficiency_ss}(A), where we show the steady state values of $P_\mathrm{long}$ as a function of $h$ and the geometric ratio $\ell_\mathrm{e}/\ell_\mathrm{m}$. The nonmonotonicity as a function of these two quantities can be understood from the nonlinear form in which these quantities appear in the mobilities $M_\mathrm{ee}$ and $M_\mathrm{me}$, which moreover determine both the deterministic and stochastic behavior of the system.  Overall,  Figure~\ref{fig:efficiency_ss}(A) serves as a validation of the golden rules (i) and (ii).

From $P_\mathrm{long}(t)$, we may calculate the probability that the molecule is in the long state once it unbinds from the enzyme, which is a quality measure for the function of the enzyme. Assuming that the molecule unbinds from the enzyme at a rate $k_\mathrm{off}$, the probability of the molecule remaining bound at time $t$ is $p_\mathrm{bo}(t)=\exp(-k_\mathrm{off}t)$. The probability of unbinding precisely between $t$ and $t+dt$ is then $p_\mathrm{bo}(t) k_\mathrm{off} dt$, and thus the average probability of being in the long state when it unbinds is
\begin{align}
    \langle P_\mathrm{long} \rangle  = k_\mathrm{off}\int^\infty_0 dt' P_\mathrm{long}(t') p_\mathrm{bo}(t') . \label{eq:avPlong}
\end{align}
The behaviour of $P_\mathrm{long}(t)$ in Figure~\ref{fig:plong_plots} can be well approximated by an exponential relaxation $P_\mathrm{long}(t) \simeq P_\mathrm{long}(\infty) - [P_\mathrm{long}(\infty)-P_\mathrm{long}(0)] \exp(-t/\tau_\mathrm{ss})$, where $\tau_\mathrm{ss}$ is the  timescale associated with the transient regime in Figure \ref{fig:plong_plots}. With this choice, Equation~\eqref{eq:avPlong} evaluates to
\begin{align}
    \langle P_\mathrm{long} \rangle  \simeq \frac{k_\mathrm{off} P_\mathrm{long}(0) +  P_\mathrm{long}(\infty) /\tau_\mathrm{ss}}{k_\mathrm{off}+1/\tau_\mathrm{ss}}.  \label{eq:avPlong2}
\end{align}
In the limiting case in which $k_\mathrm{off} \tau_\mathrm{ss} \ll 1$,  so that the typical timescale for unbinding is longer than the timescale of the transient regime in Figure~\ref{fig:plong_plots}, we find $\langle P_\mathrm{long} \rangle \simeq  P_\mathrm{long}(\infty)$.

We can also consider the energetic efficiency of the enzymatic action. In the NESS, the reaction rate of the enzyme and thus the energy dissipation (or equivalently the entropy production) rate can be used as a measure of the enzymatic activity. The enzymatic reaction rate $\Omega_\ee$ is given by
\begin{align}\label{eq:omega1}
    \Omega_\mathrm{e} \equiv \frac{1}{2\pi}\int_{-\infty}^\infty J_\mathrm{e} d\phi_\mm
\end{align}
and is constant in the NESS. The rate of energy dissipation is then given by $\Omega_\mathrm{e} E_*$. In Figure~\ref{fig:efficiency_ss}(B), we show the energy dissipation rate $\tilde{\Omega}_\mathrm{e} E_*/k_{\rm B} T$ (where $\tilde{\Omega}_\mathrm{e}$ is the rate in dimensionless time units $\tilde{t}$ as introduced above), as a function of the coupling $h$ and the geometric parameter $\ell_\ee/\ell_\mm$. Interestingly, there are large regions of parameter space where the enzyme is highly functional (larger $P_\mathrm{long}$) while the energy dissipation is much lower than the corresponding value for an uncoupled enzyme, which can be explained by a slow-down of the fuelled enzymatic reaction (smaller $\Omega_\ee$) due to the coupling to the molecular reaction.

\subsubsection{First passage time: speeding up a slow reaction}

We now focus on the early-time dynamics of the process and quantify the characteristic time that is needed for the substrate-to-product conversion to take place, i.e.~the mean first passage time. We focus on the switch reaction, although analogous results are obtained for the dissociation reaction.
We initialize the system in the short state ($\phi_\mm = -1$) at $t=0$, and numerically solve the Langevin dynamics (Equation~\eqref{eq:langevin}, see Methods for details of the numerical solution) until the long state ($\phi_\mm = 1$) is reached at a time $t=\tau_\mathrm{fpt}$ (which we then record). The ensemble average of $\tau_\mathrm{fpt}$ (i.e. average over many simulations), which we denote as $\langle \tau_\mathrm{fpt} \rangle $, corresponds to the mean first passage time.

\begin{figure*}[ht]
\centering
\includegraphics[width=0.95 \linewidth]{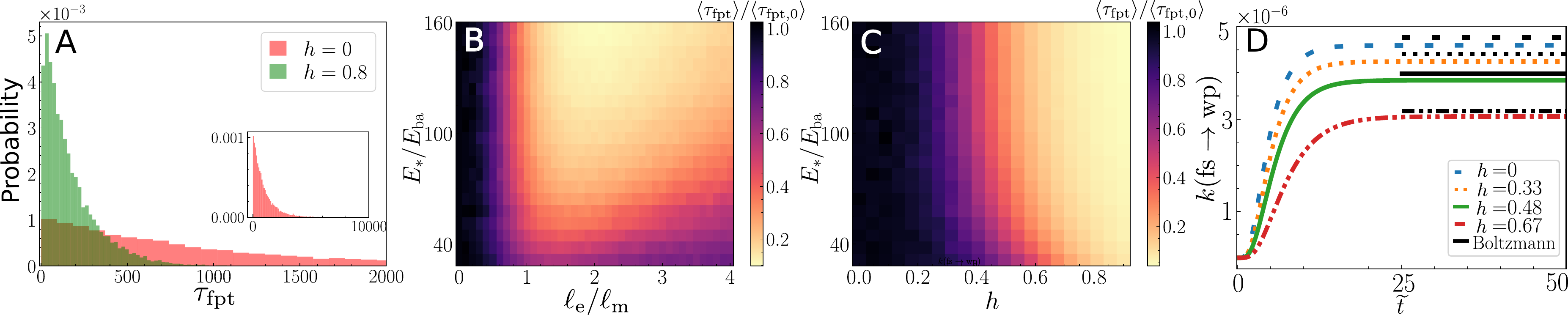}
\caption{\textbf{Results on first passage times} (A) Distribution of first passage times $\tau_\mathrm{fpt}$ for $\ell_\ee/\ell_\mm = 1.6$ in the absence (red) and presence (green) of dissipative coupling. (B,C) Mean first passage time $\langle \tau_\mathrm{fpt} \rangle$, normalized by the mean first passage time in the uncoupled case where $\ell_\ee/\ell_\mm =0$, as a function of the length ratio $\ell_\ee/\ell_\mm$ and the energy ratio $E_*/E_\mathrm{ba}$ for fixed value of coupling strength $h=0.48$ (A), and as a function of the coupling strength $h$ and $E_*/E_\mathrm{ba}$ for fixed $\ell_\ee/\ell_\mm = 1$ in (B). In (A--C), the noise strength is $k_{\rm B}T /\Delta = 0.15$. (D) The rate at which the saddle point marked as the red diamond in Figure~\ref{fig:deterministic_dynamics} is crossed extracted numerically (coloured lines) compared to the Langer rate given by Equation~\eqref{eq:langers} (black horizontal lines), for $E_\mathrm{ba}/\Delta = 0.8$ and $k_{\rm B} T/\Delta=0.08$.}\label{fig:histograms}
\end{figure*}

In Figure~\ref{fig:histograms}(A), we show the distribution of first passage times $P(\tau_\mathrm{fpt})$ for zero and large coupling strength $h$.
Clearly, in the presence of dissipative coupling, the first passage times are much smaller in relative terms, while the distribution appears to be markedly narrower. Figures \ref{fig:histograms}(B,C) show how the mean first passage time $\langle \tau_\mathrm{fpt} \rangle$, relative to the value $\langle \tau_\mathrm{fpt,0} \rangle$ corresponding to the absence of enzyme conformational changes ($\ell_\ee/\ell_\mm = 0$), depends on the tuning parameters, namely the fuelled reaction drive $E_*/E_\mathrm{ba}$, the (geometric) deformation ratio $\ell_\ee/\ell_\mm$, and the coupling strength $h$. By comparing Figure~\ref{fig:histograms}(B) with Figure~\ref{fig:parspace_diagrams}(A), it becomes clear that the speed-up in the reaction is directly related to the global bifurcation in the deterministic dynamics.

Some analytical progress towards a calculation of the mean first passage time is possible in the limit of small noise ($k_{\rm B} T$ much  smaller than the energy barriers $E_\mathrm{ba}$ and $\Delta$). In this limit, a generalization of the Kramers escape rate to higher dimensions due to Langer \cite{hanggi1990reaction,langer1968theory,langer1969statistical} gives the rate at which the probability current crosses through a saddle point, which represents the transition state. To deal with the multiplicative nature of the noise, we evaluate the mobility tensor $M_{\alpha \beta}$ around the saddle point of interest. Moreover, we are interested in the catalyzed reaction, and thus focus on the transition state shown as the red diamond in Figure~\ref{fig:deterministic_dynamics}.
The rate then takes a similar form to the one-dimensional case (see Equation~\eqref{eq:kramers}), given by 
\begin{align}\label{eq:langers}
    k(\mathrm{fs} \to \mathrm{wp}) = \frac{\vert\Lambda_-\vert}{2 \pi} \sqrt{\frac{\lambda_\mathrm{e}^{(0)}}{\vert\lambda_\mathrm{e}^{(1)}\vert}} \exp\bigg({-\frac{E_\mathrm{ba}}{k_{\rm B} T} }\bigg)
\end{align}
where $\lambda_\mathrm{e}^{(0)}=V''_\mathrm{e}(\phi_\mathrm{e})|_{\rm min}$, $\lambda_\mathrm{e}^{(1)}=V''_\mathrm{e}(\phi_\mathrm{e})|_{\rm saddle}$, and $\Lambda_-$ along with additional details is given in the Supplemental Information.
As shown in Figure~\ref{fig:histograms}(D), Equation~\eqref{eq:langers} agrees well with numerical solution of Equation~\eqref{eq:fokker-planck} for the crossing rate over the saddle point marked as the red diamond in Figure~\ref{fig:deterministic_dynamics}, after a brief initial transient. In the numerics, the initial probability distribution for both enzyme and molecule was chosen as a narrow Gaussian centered at the potential minimum. 

\section{Discussion}

We have developed a minimal model for a fuelled enzyme that is able to extract energy from a thermodynamically favourable reaction to drive and speed-up a thermodynamically unfavourable reaction.
Hence, the enzyme acts as a ``Maxwell's demon''.
The transduction is mechanical, arising from the conformational changes of the enzyme that couple to the conformational changes of the attached substrate molecule. This goes beyond previously developed minimal models for enzymes that only described passive enzymes \cite{zeravcic2014self,zeravcic2017spontaneous,zeravcic2017colloquium,rivoire2020geometry,munoz2023computational,rivoire2023flexibility}, which can speed up energetically favourable reactions but are unable to alter the reaction equilibrium as determined by thermodynamics. 

The function of our model enzyme is most prominently dictated by the geometry of the enzyme-substrate complex, and can be optimized by following three simple golden rules:
(i) the enzyme and the molecule should be attached at the smaller end of each;
(ii) the conformational change of the enzyme must be comparable to or larger than the conformational change required of the molecule;
(iii) the conformational change of the enzyme must be fast enough. These parameters can be experimentally tuned when designing an artificial enzyme, and should be experimentally accessible when investigating the function of biological enzymes. Experimental and computational studies have shown that conformational changes of different enzymes range from angstroms \cite{Gutteridge2005Feb,Lu2014Jan,Chong2024Nov} to nanometers as in the case of  hexokinase and ATPase \cite{glowacki2012taking, callender2015dynamical,Bennett1978Oct,Kuser2008Aug}. Moreover, measurements of protein binding pockets estimate the average size of the the binding pocket of enzymes to be of the order \SI{1}{\AA}--\SI{1}{\nano\meter} \cite{Liang2008Dec}. As the corresponding molecules are similar or smaller in size than the binding pocket, it appears that rule (ii) broadly applies for typical enzymes.
We note that the present model can minimally describe the slow reaction coordinate dynamics of at least four of the six main classes of enzymes \cite{stryer}. In particular, our shape-switch reaction corresponds to the action of isomerases, whereas our dissociation potential corresponds to the action of lyases. Assuming that two separate molecules can bind to the enzyme, the shape-switch reaction run in reverse would correspond to the joining action of ligases, and with minor modifications could be repurposed to describe the action of transferases, where part of one molecule is transferred to the other molecule. The remaining two main classes of enzymes, namely, oxidoreductases and hydrolases, perform highly chemically specific actions that may not be amenable to a purely mechanical description.

In general, we were able to understand the emergence of a strong coupling between the thermodynamically favourable and unfavourable reactions to be a result of a global bifurcation in the deterministic dynamics of the system, of a similar nature to that causing synchronization and phase-locking among mechanically-coupled enzymes \cite{jaime,chatzittofi2023topological}, which can explain the enhanced activity of enzyme oligomeric complexes as observed in some experiments \cite{Lu2014,Hehir2000}.
Importantly, by adding thermodynamically-consistent fluctuations to the theory, we showed that the effect survives in the presence of noise, and we could calculate important quantities such as non-equilibrium steady states reflecting the departure from equilibrium thermodynamics in the reaction equilibrium, and first passage time distributions for the catalyzed reaction. Interestingly, the effectiveness of the enzyme (as measured by successful catalytic action) and its energy dissipation do not follow a one-to-one relation (Figure~\ref{fig:efficiency_ss}), leaving additional room for the optimization of not just the enzyme effectiveness but its energetic efficiency. 

Our approach readily allows us to probe the energetic aspects of the non-equilibrium reaction \cite{Ragazzon2018}, and in particular, the entropy production due to the coupling between the mechanical and the chemical degrees of freedom \cite{chatzittofi2023entropy}. 
The proposed design rules in this work focused on optimizing the transition step of the reaction, assuming that the binding of the substrate and unbinding of the product from the enzyme is essentially instantaneous and perfectly specific, and that the enzyme-substrate bond is strong enough to avoid unbinding of the substrate during the (incomplete) reaction. A future and more detailed description of these binding and unbinding processes is an obvious next step for the further optimization of the enzymatic activity, now including selectivity. In any case, the present mechanistic framework (and the resulting rules) can provide physical intuition to be used as input to complement atomistic simulation and machine learning approaches for de novo design of proteins with dynamic function such as enzymes \cite{Frank2024Oct}. It can also be used to study the evolutionary advantage of having mechanical control during the catalytic process via conformational changes, as opposed to simple utilization of uncontrolled metal-based catalysis.

In conclusion, our model introduces new and important features to serve as a minimal model of a fuelled catalyst. This opens an avenue towards further progress in understanding and designing bio-inspired systems for artificial catalysis. 

\acknowledgements
We acknowledge discussions with Navdeep Rana regarding the numerical integrators, and support from the Max Planck School Matter to Life and the MaxSynBio Consortium which are jointly funded by the Federal Ministry of Education and Research (BMBF) of Germany and the Max Planck Society.

\section*{Methods}

\subsection*{Simulations}\label{methods}

To generate the phase portraits in Figure~\ref{fig:deterministic_dynamics} and the parameter scan in Figure~\ref{fig:parspace_diagrams} from the deterministic equations (Equation~\eqref{eq:deterministic}), we used the built-in \textit{ode45} integrator of MATLAB, which uses a fourth order Runge-Kutta method \cite{MATLAB}. A grid of $201 \times 201$ for $-\pi<\phi_\ee<\pi$ and $-2<\phi_\mm<2$ was used to generate the initial conditions, and the equations were integrated for a total time of $\tilde t_{\rm tot} = 120$. 

For the numerical integration of the Fokker-Planck equation given by Equation~\eqref{eq:fokker-planck} we used a custom code written in Python \cite{python}. The code uses a finite difference method with fourth-order accuracy for the spatial derivatives and fourth order Runge-Kutta method for the time integration. For the enzyme coordinate $\phi_\ee$ we considered the domain $0 \leq \phi_\ee < 2\pi$ with periodic boundary conditions. For the molecule coordinate $\phi_\mm$, we used Dirichlet boundary conditions in the domains $[-1.8,1.8],[-2.0,2.0],[-2.2,2.2]$ and noise strengths $k_{\rm B} T/\Delta = 0.15,0.2,0.4$.
To discretize the domains we used a rectangular shape domain such that the finite-difference increments satisfied $\Delta \phi_\ee \simeq \Delta \phi_\mm$ and thus the integrator would be more stable. For $k_{\rm B} T/\Delta = 0.15,0.2,0.4$ the corresponding number of points $(N_1,N_2)$ per side were $(512,128), (400,128), (300,100)$, and the time step was $\Delta \tilde t = 10^{-3}, 10^{-3}, 5 \cdot 10^{-4}$.

For the numerical integration of the Langevin equations in Equation~\eqref{eq:langevin}, we used a custom code written in Julia \cite{julia}. The Euler-Maruyama method was used for the numerical integration. The time step used was $\Delta \tilde t = 10^{-3}$.

% \begin{figure}[t]
% \centering
% \includegraphics[width=1. \linewidth]{mirror-tog.pdf}
% \caption{\textbf{Reverse switch reaction.} (A) Phase-portrait for $h=0.48$, $\ell_\ee/\ell_\mm = 2.5$, and $\varepsilon/\Delta =-0.1$. The black solid line represents a deterministic trajectory. (B) Marginal distributions of the molecule reaction coordinate in steady-state, for various values of the coupling strength. (A) and (B) can be respectively compared to Figure~\ref{fig:deterministic_dynamics}(D) and Figure~\ref{fig:ss_distributions}(B) for the forward reaction.}\label{fig:mirror}
% \end{figure}

To check for consistency between the Langevin and Fokker-Planck approaches, long trajectories from the solution of the Langevin equations were binned and compared with the steady state distribution predicted by the Fokker-Planck equation. The two approaches exhibit good agreement, as depicted in Figure~S2 for the case of $h=0.67$ and $k_{\rm B}T/\Delta =0.2$.

\bibliography{bibtex}

\providecommand{\noopsort}[1]{}\providecommand{\singleletter}[1]{#1}%
\begin{thebibliography}{73}%
\makeatletter
\providecommand \@ifxundefined [1]{%
 \@ifx{#1\undefined}
}%
\providecommand \@ifnum [1]{%
 \ifnum #1\expandafter \@firstoftwo
 \else \expandafter \@secondoftwo
 \fi
}%
\providecommand \@ifx [1]{%
 \ifx #1\expandafter \@firstoftwo
 \else \expandafter \@secondoftwo
 \fi
}%
\providecommand \natexlab [1]{#1}%
\providecommand \enquote  [1]{``#1''}%
\providecommand \bibnamefont  [1]{#1}%
\providecommand \bibfnamefont [1]{#1}%
\providecommand \citenamefont [1]{#1}%
\providecommand \href@noop [0]{\@secondoftwo}%
\providecommand \href [0]{\begingroup \@sanitize@url \@href}%
\providecommand \@href[1]{\@@startlink{#1}\@@href}%
\providecommand \@@href[1]{\endgroup#1\@@endlink}%
\providecommand \@sanitize@url [0]{\catcode `\\12\catcode `\$12\catcode
  `\&12\catcode `\#12\catcode `\^12\catcode `\_12\catcode `\%12\relax}%
\providecommand \@@startlink[1]{}%
\providecommand \@@endlink[0]{}%
\providecommand \url  [0]{\begingroup\@sanitize@url \@url }%
\providecommand \@url [1]{\endgroup\@href {#1}{\urlprefix }}%
\providecommand \urlprefix  [0]{URL }%
\providecommand \Eprint [0]{\href }%
\providecommand \doibase [0]{https://doi.org/}%
\providecommand \selectlanguage [0]{\@gobble}%
\providecommand \bibinfo  [0]{\@secondoftwo}%
\providecommand \bibfield  [0]{\@secondoftwo}%
\providecommand \translation [1]{[#1]}%
\providecommand \BibitemOpen [0]{}%
\providecommand \bibitemStop [0]{}%
\providecommand \bibitemNoStop [0]{.\EOS\space}%
\providecommand \EOS [0]{\spacefactor3000\relax}%
\providecommand \BibitemShut  [1]{\csname bibitem#1\endcsname}%
\let\auto@bib@innerbib\@empty
%</preamble>
\bibitem [{\citenamefont {Borsley}\ \emph {et~al.}(2024)\citenamefont
  {Borsley}, \citenamefont {Leigh},\ and\ \citenamefont
  {Roberts}}]{Borsley2024}%
  \BibitemOpen
  \bibfield  {author} {\bibinfo {author} {\bibfnamefont {S.}~\bibnamefont
  {Borsley}}, \bibinfo {author} {\bibfnamefont {D.~A.}\ \bibnamefont {Leigh}},\
  and\ \bibinfo {author} {\bibfnamefont {B.~M.~W.}\ \bibnamefont {Roberts}},\
  }\bibfield  {title} {\bibinfo {title} {{Molecular Ratchets and Kinetic
  Asymmetry: Giving Chemistry Direction}},\ }\href
  {https://doi.org/10.1002/anie.202400495} {\bibfield  {journal} {\bibinfo
  {journal} {Angew. Chem. Int. Ed.}\ }\textbf {\bibinfo {volume} {63}},\
  \bibinfo {pages} {e202400495} (\bibinfo {year} {2024})}\BibitemShut {NoStop}%
\bibitem [{\citenamefont {Michaelis}\ and\ \citenamefont
  {Menten}(1913)}]{MichaelisMenten}%
  \BibitemOpen
  \bibfield  {author} {\bibinfo {author} {\bibfnamefont {L.}~\bibnamefont
  {Michaelis}}\ and\ \bibinfo {author} {\bibfnamefont {M.~L.}\ \bibnamefont
  {Menten}},\ }\bibfield  {title} {\bibinfo {title} {Die kinetik der
  invertinwirkung},\ }\href@noop {} {\bibfield  {journal} {\bibinfo  {journal}
  {Biochemische Zeitschrift}\ }\textbf {\bibinfo {volume} {49}},\ \bibinfo
  {pages} {333} (\bibinfo {year} {1913})}\BibitemShut {NoStop}%
\bibitem [{\citenamefont {Johnson}\ and\ \citenamefont
  {Goody}(2011)}]{johnson2011original}%
  \BibitemOpen
  \bibfield  {author} {\bibinfo {author} {\bibfnamefont {K.~A.}\ \bibnamefont
  {Johnson}}\ and\ \bibinfo {author} {\bibfnamefont {R.~S.}\ \bibnamefont
  {Goody}},\ }\bibfield  {title} {\bibinfo {title} {{The Original Michaelis
  Constant: Translation of the 1913 Michaelis{\textendash}Menten Paper}},\
  }\href {https://doi.org/10.1021/bi201284u} {\bibfield  {journal} {\bibinfo
  {journal} {Biochemistry}\ }\textbf {\bibinfo {volume} {50}},\ \bibinfo
  {pages} {8264} (\bibinfo {year} {2011})}\BibitemShut {NoStop}%
\bibitem [{\citenamefont {Qian}(2007)}]{qian2007phosphorylation}%
  \BibitemOpen
  \bibfield  {author} {\bibinfo {author} {\bibfnamefont {H.}~\bibnamefont
  {Qian}},\ }\bibfield  {title} {\bibinfo {title} {Phosphorylation energy
  hypothesis: Open chemical systems and their biological functions},\ }\href
  {https://doi.org/10.1146/annurev.physchem.58.032806.104550} {\bibfield
  {journal} {\bibinfo  {journal} {Annu. Rev. Phys. Chem.}\ }\textbf {\bibinfo
  {volume} {58}},\ \bibinfo {pages} {113} (\bibinfo {year} {2007})}\BibitemShut
  {NoStop}%
\bibitem [{\citenamefont {Lu}(1998)}]{Lu1998}%
  \BibitemOpen
  \bibfield  {author} {\bibinfo {author} {\bibfnamefont {H.~P.}\ \bibnamefont
  {Lu}},\ }\bibfield  {title} {\bibinfo {title} {Single-molecule enzymatic
  dynamics},\ }\href {https://doi.org/10.1126/science.282.5395.1877} {\bibfield
   {journal} {\bibinfo  {journal} {Science}\ }\textbf {\bibinfo {volume}
  {282}},\ \bibinfo {pages} {1877–1882} (\bibinfo {year} {1998})}\BibitemShut
  {NoStop}%
\bibitem [{\citenamefont {English}\ \emph {et~al.}(2005)\citenamefont
  {English}, \citenamefont {Min}, \citenamefont {van Oijen}, \citenamefont
  {Lee}, \citenamefont {Luo}, \citenamefont {Sun}, \citenamefont {Cherayil},
  \citenamefont {Kou},\ and\ \citenamefont {Xie}}]{English2005}%
  \BibitemOpen
  \bibfield  {author} {\bibinfo {author} {\bibfnamefont {B.~P.}\ \bibnamefont
  {English}}, \bibinfo {author} {\bibfnamefont {W.}~\bibnamefont {Min}},
  \bibinfo {author} {\bibfnamefont {A.~M.}\ \bibnamefont {van Oijen}}, \bibinfo
  {author} {\bibfnamefont {K.~T.}\ \bibnamefont {Lee}}, \bibinfo {author}
  {\bibfnamefont {G.}~\bibnamefont {Luo}}, \bibinfo {author} {\bibfnamefont
  {H.}~\bibnamefont {Sun}}, \bibinfo {author} {\bibfnamefont {B.~J.}\
  \bibnamefont {Cherayil}}, \bibinfo {author} {\bibfnamefont {S.~C.}\
  \bibnamefont {Kou}},\ and\ \bibinfo {author} {\bibfnamefont {X.~S.}\
  \bibnamefont {Xie}},\ }\bibfield  {title} {\bibinfo {title}
  {{Ever-fluctuating single enzyme molecules: Michaelis-Menten equation
  revisited}},\ }\href {https://doi.org/10.1038/nchembio759} {\bibfield
  {journal} {\bibinfo  {journal} {Nat. Chem. Biol.}\ }\textbf {\bibinfo
  {volume} {2}},\ \bibinfo {pages} {87–94} (\bibinfo {year}
  {2005})}\BibitemShut {NoStop}%
\bibitem [{\citenamefont {Kou}\ \emph {et~al.}(2005)\citenamefont {Kou},
  \citenamefont {Cherayil}, \citenamefont {Min}, \citenamefont {English},\ and\
  \citenamefont {Xie}}]{Kou2005}%
  \BibitemOpen
  \bibfield  {author} {\bibinfo {author} {\bibfnamefont {S.}~\bibnamefont
  {Kou}}, \bibinfo {author} {\bibfnamefont {B.~J.}\ \bibnamefont {Cherayil}},
  \bibinfo {author} {\bibfnamefont {W.}~\bibnamefont {Min}}, \bibinfo {author}
  {\bibfnamefont {B.~P.}\ \bibnamefont {English}},\ and\ \bibinfo {author}
  {\bibfnamefont {X.~S.}\ \bibnamefont {Xie}},\ }\bibfield  {title} {\bibinfo
  {title} {Single-molecule michaelis- menten equations},\ }\href
  {https://doi.org/10.1021/jp051490q} {\bibfield  {journal} {\bibinfo
  {journal} {J. Phys. Chem. B}\ }\textbf {\bibinfo {volume} {109}},\ \bibinfo
  {pages} {19068} (\bibinfo {year} {2005})}\BibitemShut {NoStop}%
\bibitem [{\citenamefont {Kramers}(1940)}]{kramers1940brownian}%
  \BibitemOpen
  \bibfield  {author} {\bibinfo {author} {\bibfnamefont {H.}~\bibnamefont
  {Kramers}},\ }\bibfield  {title} {\bibinfo {title} {Brownian motion in a
  field of force and the diffusion model of chemical reactions},\ }\href
  {https://doi.org/10.1016/s0031-8914(40)90098-2} {\bibfield  {journal}
  {\bibinfo  {journal} {Physica}\ }\textbf {\bibinfo {volume} {7}},\ \bibinfo
  {pages} {284} (\bibinfo {year} {1940})}\BibitemShut {NoStop}%
\bibitem [{\citenamefont {H\"anggi}\ \emph {et~al.}(1990)\citenamefont
  {H\"anggi}, \citenamefont {Talkner},\ and\ \citenamefont
  {Borkovec}}]{hanggi1990reaction}%
  \BibitemOpen
  \bibfield  {author} {\bibinfo {author} {\bibfnamefont {P.}~\bibnamefont
  {H\"anggi}}, \bibinfo {author} {\bibfnamefont {P.}~\bibnamefont {Talkner}},\
  and\ \bibinfo {author} {\bibfnamefont {M.}~\bibnamefont {Borkovec}},\
  }\bibfield  {title} {\bibinfo {title} {{Reaction-rate theory: fifty years
  after Kramers}},\ }\href {https://doi.org/10.1103/RevModPhys.62.251}
  {\bibfield  {journal} {\bibinfo  {journal} {Rev. Mod. Phys.}\ }\textbf
  {\bibinfo {volume} {62}},\ \bibinfo {pages} {251} (\bibinfo {year}
  {1990})}\BibitemShut {NoStop}%
\bibitem [{\citenamefont {Karplus}\ and\ \citenamefont
  {McCammon}(2002)}]{Karplus2002}%
  \BibitemOpen
  \bibfield  {author} {\bibinfo {author} {\bibfnamefont {M.}~\bibnamefont
  {Karplus}}\ and\ \bibinfo {author} {\bibfnamefont {J.~A.}\ \bibnamefont
  {McCammon}},\ }\bibfield  {title} {\bibinfo {title} {Molecular dynamics
  simulations of biomolecules},\ }\href {https://doi.org/10.1038/nsb0902-646}
  {\bibfield  {journal} {\bibinfo  {journal} {Nat. Struct. Mol. Biol.}\
  }\textbf {\bibinfo {volume} {9}},\ \bibinfo {pages} {646} (\bibinfo {year}
  {2002})}\BibitemShut {NoStop}%
\bibitem [{\citenamefont {Kutzner}\ \emph {et~al.}(2019)\citenamefont
  {Kutzner}, \citenamefont {Páll}, \citenamefont {Fechner}, \citenamefont
  {Esztermann}, \citenamefont {de~Groot},\ and\ \citenamefont
  {Grubm\"{u}ller}}]{Kutzner2019}%
  \BibitemOpen
  \bibfield  {author} {\bibinfo {author} {\bibfnamefont {C.}~\bibnamefont
  {Kutzner}}, \bibinfo {author} {\bibfnamefont {S.}~\bibnamefont {Páll}},
  \bibinfo {author} {\bibfnamefont {M.}~\bibnamefont {Fechner}}, \bibinfo
  {author} {\bibfnamefont {A.}~\bibnamefont {Esztermann}}, \bibinfo {author}
  {\bibfnamefont {B.~L.}\ \bibnamefont {de~Groot}},\ and\ \bibinfo {author}
  {\bibfnamefont {H.}~\bibnamefont {Grubm\"{u}ller}},\ }\bibfield  {title}
  {\bibinfo {title} {{More bang for your buck: Improved use of GPU nodes for
  GROMACS 2018}},\ }\href {https://doi.org/10.1002/jcc.26011} {\bibfield
  {journal} {\bibinfo  {journal} {J. Comput. Chem.}\ }\textbf {\bibinfo
  {volume} {40}},\ \bibinfo {pages} {2418–2431} (\bibinfo {year}
  {2019})}\BibitemShut {NoStop}%
\bibitem [{\citenamefont {Lovelock}\ \emph {et~al.}(2022)\citenamefont
  {Lovelock}, \citenamefont {Crawshaw}, \citenamefont {Basler}, \citenamefont
  {Levy}, \citenamefont {Baker}, \citenamefont {Hilvert},\ and\ \citenamefont
  {Green}}]{Lovelock2022}%
  \BibitemOpen
  \bibfield  {author} {\bibinfo {author} {\bibfnamefont {S.~L.}\ \bibnamefont
  {Lovelock}}, \bibinfo {author} {\bibfnamefont {R.}~\bibnamefont {Crawshaw}},
  \bibinfo {author} {\bibfnamefont {S.}~\bibnamefont {Basler}}, \bibinfo
  {author} {\bibfnamefont {C.}~\bibnamefont {Levy}}, \bibinfo {author}
  {\bibfnamefont {D.}~\bibnamefont {Baker}}, \bibinfo {author} {\bibfnamefont
  {D.}~\bibnamefont {Hilvert}},\ and\ \bibinfo {author} {\bibfnamefont {A.~P.}\
  \bibnamefont {Green}},\ }\bibfield  {title} {\bibinfo {title} {The road to
  fully programmable protein catalysis},\ }\href
  {https://doi.org/10.1038/s41586-022-04456-z} {\bibfield  {journal} {\bibinfo
  {journal} {Nature}\ }\textbf {\bibinfo {volume} {606}},\ \bibinfo {pages}
  {49} (\bibinfo {year} {2022})}\BibitemShut {NoStop}%
\bibitem [{\citenamefont {Yeh}\ \emph {et~al.}(2023)\citenamefont {Yeh},
  \citenamefont {Norn}, \citenamefont {Kipnis}, \citenamefont {Tischer},
  \citenamefont {Pellock}, \citenamefont {Evans}, \citenamefont {Ma},
  \citenamefont {Lee}, \citenamefont {Zhang}, \citenamefont {Anishchenko},
  \citenamefont {Coventry}, \citenamefont {Cao}, \citenamefont {Dauparas},
  \citenamefont {Halabiya}, \citenamefont {DeWitt}, \citenamefont {Carter},
  \citenamefont {Houk},\ and\ \citenamefont {Baker}}]{Yeh2023}%
  \BibitemOpen
  \bibfield  {author} {\bibinfo {author} {\bibfnamefont {A.~H.-W.}\
  \bibnamefont {Yeh}}, \bibinfo {author} {\bibfnamefont {C.}~\bibnamefont
  {Norn}}, \bibinfo {author} {\bibfnamefont {Y.}~\bibnamefont {Kipnis}},
  \bibinfo {author} {\bibfnamefont {D.}~\bibnamefont {Tischer}}, \bibinfo
  {author} {\bibfnamefont {S.~J.}\ \bibnamefont {Pellock}}, \bibinfo {author}
  {\bibfnamefont {D.}~\bibnamefont {Evans}}, \bibinfo {author} {\bibfnamefont
  {P.}~\bibnamefont {Ma}}, \bibinfo {author} {\bibfnamefont {G.~R.}\
  \bibnamefont {Lee}}, \bibinfo {author} {\bibfnamefont {J.~Z.}\ \bibnamefont
  {Zhang}}, \bibinfo {author} {\bibfnamefont {I.}~\bibnamefont {Anishchenko}},
  \bibinfo {author} {\bibfnamefont {B.}~\bibnamefont {Coventry}}, \bibinfo
  {author} {\bibfnamefont {L.}~\bibnamefont {Cao}}, \bibinfo {author}
  {\bibfnamefont {J.}~\bibnamefont {Dauparas}}, \bibinfo {author}
  {\bibfnamefont {S.}~\bibnamefont {Halabiya}}, \bibinfo {author}
  {\bibfnamefont {M.}~\bibnamefont {DeWitt}}, \bibinfo {author} {\bibfnamefont
  {L.}~\bibnamefont {Carter}}, \bibinfo {author} {\bibfnamefont {K.~N.}\
  \bibnamefont {Houk}},\ and\ \bibinfo {author} {\bibfnamefont
  {D.}~\bibnamefont {Baker}},\ }\bibfield  {title} {\bibinfo {title} {De novo
  design of luciferases using deep learning},\ }\href
  {https://doi.org/10.1038/s41586-023-05696-3} {\bibfield  {journal} {\bibinfo
  {journal} {Nature}\ }\textbf {\bibinfo {volume} {614}},\ \bibinfo {pages}
  {774–780} (\bibinfo {year} {2023})}\BibitemShut {NoStop}%
\bibitem [{\citenamefont {Dallago}\ and\ \citenamefont
  {Yang}(2023)}]{Dallago2023}%
  \BibitemOpen
  \bibfield  {author} {\bibinfo {author} {\bibfnamefont {C.}~\bibnamefont
  {Dallago}}\ and\ \bibinfo {author} {\bibfnamefont {K.~K.}\ \bibnamefont
  {Yang}},\ }\bibfield  {title} {\bibinfo {title} {Illuminating enzyme design
  using deep learning},\ }\href {https://doi.org/10.1038/s41557-023-01218-w}
  {\bibfield  {journal} {\bibinfo  {journal} {Nat. Chem.}\ }\textbf {\bibinfo
  {volume} {15}},\ \bibinfo {pages} {749–750} (\bibinfo {year}
  {2023})}\BibitemShut {NoStop}%
\bibitem [{\citenamefont {Kathan}\ \emph {et~al.}(2021)\citenamefont {Kathan},
  \citenamefont {Crespi}, \citenamefont {Thiel}, \citenamefont {Stares},
  \citenamefont {Morsa}, \citenamefont {de~Boer}, \citenamefont {Pacella},
  \citenamefont {van~den Enk}, \citenamefont {Kobauri}, \citenamefont
  {Portale}, \citenamefont {Schalley},\ and\ \citenamefont
  {Feringa}}]{Kathan2021}%
  \BibitemOpen
  \bibfield  {author} {\bibinfo {author} {\bibfnamefont {M.}~\bibnamefont
  {Kathan}}, \bibinfo {author} {\bibfnamefont {S.}~\bibnamefont {Crespi}},
  \bibinfo {author} {\bibfnamefont {N.~O.}\ \bibnamefont {Thiel}}, \bibinfo
  {author} {\bibfnamefont {D.~L.}\ \bibnamefont {Stares}}, \bibinfo {author}
  {\bibfnamefont {D.}~\bibnamefont {Morsa}}, \bibinfo {author} {\bibfnamefont
  {J.}~\bibnamefont {de~Boer}}, \bibinfo {author} {\bibfnamefont
  {G.}~\bibnamefont {Pacella}}, \bibinfo {author} {\bibfnamefont
  {T.}~\bibnamefont {van~den Enk}}, \bibinfo {author} {\bibfnamefont
  {P.}~\bibnamefont {Kobauri}}, \bibinfo {author} {\bibfnamefont
  {G.}~\bibnamefont {Portale}}, \bibinfo {author} {\bibfnamefont {C.~A.}\
  \bibnamefont {Schalley}},\ and\ \bibinfo {author} {\bibfnamefont {B.~L.}\
  \bibnamefont {Feringa}},\ }\bibfield  {title} {\bibinfo {title} {A
  light-fuelled nanoratchet shifts a coupled chemical equilibrium},\ }\href
  {https://doi.org/10.1038/s41565-021-01021-z} {\bibfield  {journal} {\bibinfo
  {journal} {Nat. Nanotechnol.}\ }\textbf {\bibinfo {volume} {17}},\ \bibinfo
  {pages} {159–165} (\bibinfo {year} {2021})}\BibitemShut {NoStop}%
\bibitem [{\citenamefont {Corra}\ \emph {et~al.}(2022)\citenamefont {Corra},
  \citenamefont {Bakić}, \citenamefont {Groppi}, \citenamefont {Baroncini},
  \citenamefont {Silvi}, \citenamefont {Penocchio}, \citenamefont {Esposito},\
  and\ \citenamefont {Credi}}]{Corra2022}%
  \BibitemOpen
  \bibfield  {author} {\bibinfo {author} {\bibfnamefont {S.}~\bibnamefont
  {Corra}}, \bibinfo {author} {\bibfnamefont {M.~T.}\ \bibnamefont {Bakić}},
  \bibinfo {author} {\bibfnamefont {J.}~\bibnamefont {Groppi}}, \bibinfo
  {author} {\bibfnamefont {M.}~\bibnamefont {Baroncini}}, \bibinfo {author}
  {\bibfnamefont {S.}~\bibnamefont {Silvi}}, \bibinfo {author} {\bibfnamefont
  {E.}~\bibnamefont {Penocchio}}, \bibinfo {author} {\bibfnamefont
  {M.}~\bibnamefont {Esposito}},\ and\ \bibinfo {author} {\bibfnamefont
  {A.}~\bibnamefont {Credi}},\ }\bibfield  {title} {\bibinfo {title} {Kinetic
  and energetic insights into the dissipative non-equilibrium operation of an
  autonomous light-powered supramolecular pump},\ }\href
  {https://doi.org/10.1038/s41565-022-01151-y} {\bibfield  {journal} {\bibinfo
  {journal} {Nat. Nanotechnol.}\ }\textbf {\bibinfo {volume} {17}},\ \bibinfo
  {pages} {746–751} (\bibinfo {year} {2022})}\BibitemShut {NoStop}%
\bibitem [{\citenamefont {Sorrenti}\ \emph {et~al.}(2017)\citenamefont
  {Sorrenti}, \citenamefont {Leira-Iglesias}, \citenamefont {Sato},\ and\
  \citenamefont {Hermans}}]{Sorrenti2017}%
  \BibitemOpen
  \bibfield  {author} {\bibinfo {author} {\bibfnamefont {A.}~\bibnamefont
  {Sorrenti}}, \bibinfo {author} {\bibfnamefont {J.}~\bibnamefont
  {Leira-Iglesias}}, \bibinfo {author} {\bibfnamefont {A.}~\bibnamefont
  {Sato}},\ and\ \bibinfo {author} {\bibfnamefont {T.~M.}\ \bibnamefont
  {Hermans}},\ }\bibfield  {title} {\bibinfo {title} {Non-equilibrium steady
  states in supramolecular polymerization},\ }\href
  {http://dx.doi.org/10.1038/ncomms15899} {\bibfield  {journal} {\bibinfo
  {journal} {Nat. Commun.}\ }\textbf {\bibinfo {volume} {8}} (\bibinfo {year}
  {2017})}\BibitemShut {NoStop}%
\bibitem [{\citenamefont {Pumm}\ \emph {et~al.}(2022)\citenamefont {Pumm},
  \citenamefont {Engelen}, \citenamefont {Kopperger}, \citenamefont {Isensee},
  \citenamefont {Vogt}, \citenamefont {Kozina}, \citenamefont {Kube},
  \citenamefont {Honemann}, \citenamefont {Bertosin}, \citenamefont
  {Langecker}, \citenamefont {Golestanian}, \citenamefont {Simmel},\ and\
  \citenamefont {Dietz}}]{Pumm2022}%
  \BibitemOpen
  \bibfield  {author} {\bibinfo {author} {\bibfnamefont {A.-K.}\ \bibnamefont
  {Pumm}}, \bibinfo {author} {\bibfnamefont {W.}~\bibnamefont {Engelen}},
  \bibinfo {author} {\bibfnamefont {E.}~\bibnamefont {Kopperger}}, \bibinfo
  {author} {\bibfnamefont {J.}~\bibnamefont {Isensee}}, \bibinfo {author}
  {\bibfnamefont {M.}~\bibnamefont {Vogt}}, \bibinfo {author} {\bibfnamefont
  {V.}~\bibnamefont {Kozina}}, \bibinfo {author} {\bibfnamefont
  {M.}~\bibnamefont {Kube}}, \bibinfo {author} {\bibfnamefont {M.~N.}\
  \bibnamefont {Honemann}}, \bibinfo {author} {\bibfnamefont {E.}~\bibnamefont
  {Bertosin}}, \bibinfo {author} {\bibfnamefont {M.}~\bibnamefont {Langecker}},
  \bibinfo {author} {\bibfnamefont {R.}~\bibnamefont {Golestanian}}, \bibinfo
  {author} {\bibfnamefont {F.~C.}\ \bibnamefont {Simmel}},\ and\ \bibinfo
  {author} {\bibfnamefont {H.}~\bibnamefont {Dietz}},\ }\bibfield  {title}
  {\bibinfo {title} {A {DNA} origami rotary ratchet motor},\ }\href
  {https://doi.org/10.1038/s41586-022-04910-y} {\bibfield  {journal} {\bibinfo
  {journal} {Nature}\ }\textbf {\bibinfo {volume} {607}},\ \bibinfo {pages}
  {492–498} (\bibinfo {year} {2022})}\BibitemShut {NoStop}%
\bibitem [{\citenamefont {Osat}\ \emph {et~al.}(2024)\citenamefont {Osat},
  \citenamefont {Metson}, \citenamefont {Kardar},\ and\ \citenamefont
  {Golestanian}}]{Osat2024}%
  \BibitemOpen
  \bibfield  {author} {\bibinfo {author} {\bibfnamefont {S.}~\bibnamefont
  {Osat}}, \bibinfo {author} {\bibfnamefont {J.}~\bibnamefont {Metson}},
  \bibinfo {author} {\bibfnamefont {M.}~\bibnamefont {Kardar}},\ and\ \bibinfo
  {author} {\bibfnamefont {R.}~\bibnamefont {Golestanian}},\ }\bibfield
  {title} {\bibinfo {title} {Escaping kinetic traps using nonreciprocal
  interactions},\ }\href {https://doi.org/10.1103/PhysRevLett.133.028301}
  {\bibfield  {journal} {\bibinfo  {journal} {Phys. Rev. Lett.}\ }\textbf
  {\bibinfo {volume} {133}},\ \bibinfo {pages} {028301} (\bibinfo {year}
  {2024})}\BibitemShut {NoStop}%
\bibitem [{\citenamefont {Navas}\ and\ \citenamefont
  {Klapp}(2024)}]{Klapp2024}%
  \BibitemOpen
  \bibfield  {author} {\bibinfo {author} {\bibfnamefont {S.~F.}\ \bibnamefont
  {Navas}}\ and\ \bibinfo {author} {\bibfnamefont {S.~H.~L.}\ \bibnamefont
  {Klapp}},\ }\href {https://doi.org/10.48550/ARXIV.2404.12108} {\bibinfo
  {title} {Impact of non-reciprocal interactions on colloidal self-assembly
  with tunable anisotropy}} (\bibinfo {year} {2024})\BibitemShut {NoStop}%
\bibitem [{\citenamefont {Osat}\ and\ \citenamefont
  {Golestanian}(2023)}]{Osat2023}%
  \BibitemOpen
  \bibfield  {author} {\bibinfo {author} {\bibfnamefont {S.}~\bibnamefont
  {Osat}}\ and\ \bibinfo {author} {\bibfnamefont {R.}~\bibnamefont
  {Golestanian}},\ }\bibfield  {title} {\bibinfo {title} {Non-reciprocal
  multifarious self-organization},\ }\href
  {https://doi.org/10.1038/s41565-022-01258-2} {\bibfield  {journal} {\bibinfo
  {journal} {Nat. Nanotechnol.}\ }\textbf {\bibinfo {volume} {18}},\ \bibinfo
  {pages} {79} (\bibinfo {year} {2023})}\BibitemShut {NoStop}%
\bibitem [{\citenamefont {Ouazan-Reboul}\ \emph {et~al.}(2023)\citenamefont
  {Ouazan-Reboul}, \citenamefont {Agudo-Canalejo},\ and\ \citenamefont
  {Golestanian}}]{OuazanReboul2023}%
  \BibitemOpen
  \bibfield  {author} {\bibinfo {author} {\bibfnamefont {V.}~\bibnamefont
  {Ouazan-Reboul}}, \bibinfo {author} {\bibfnamefont {J.}~\bibnamefont
  {Agudo-Canalejo}},\ and\ \bibinfo {author} {\bibfnamefont {R.}~\bibnamefont
  {Golestanian}},\ }\bibfield  {title} {\bibinfo {title} {Self-organization of
  primitive metabolic cycles due to non-reciprocal interactions},\ }\href
  {http://dx.doi.org/10.1038/s41467-023-40241-w} {\bibfield  {journal}
  {\bibinfo  {journal} {Nat. Commun.}\ }\textbf {\bibinfo {volume} {14}},\
  \bibinfo {pages} {4496} (\bibinfo {year} {2023})}\BibitemShut {NoStop}%
\bibitem [{\citenamefont {Glowacki}\ \emph {et~al.}(2012)\citenamefont
  {Glowacki}, \citenamefont {Harvey},\ and\ \citenamefont
  {Mulholland}}]{glowacki2012taking}%
  \BibitemOpen
  \bibfield  {author} {\bibinfo {author} {\bibfnamefont {D.~R.}\ \bibnamefont
  {Glowacki}}, \bibinfo {author} {\bibfnamefont {J.~N.}\ \bibnamefont
  {Harvey}},\ and\ \bibinfo {author} {\bibfnamefont {A.~J.}\ \bibnamefont
  {Mulholland}},\ }\bibfield  {title} {\bibinfo {title} {Taking
  ockham{\textquotesingle}s razor to enzyme dynamics and catalysis},\ }\href
  {https://doi.org/10.1038/nchem.1244} {\bibfield  {journal} {\bibinfo
  {journal} {Nat. Chem.}\ }\textbf {\bibinfo {volume} {4}},\ \bibinfo {pages}
  {169} (\bibinfo {year} {2012})}\BibitemShut {NoStop}%
\bibitem [{\citenamefont {Callender}\ and\ \citenamefont
  {Dyer}(2014)}]{callender2015dynamical}%
  \BibitemOpen
  \bibfield  {author} {\bibinfo {author} {\bibfnamefont {R.}~\bibnamefont
  {Callender}}\ and\ \bibinfo {author} {\bibfnamefont {R.~B.}\ \bibnamefont
  {Dyer}},\ }\bibfield  {title} {\bibinfo {title} {The dynamical nature of
  enzymatic catalysis},\ }\href {https://doi.org/10.1021/ar5002928} {\bibfield
  {journal} {\bibinfo  {journal} {Acc. Chem. Res.}\ }\textbf {\bibinfo {volume}
  {48}},\ \bibinfo {pages} {407} (\bibinfo {year} {2014})}\BibitemShut
  {NoStop}%
\bibitem [{\citenamefont {J\"ulicher}\ \emph {et~al.}(1997)\citenamefont
  {J\"ulicher}, \citenamefont {Ajdari},\ and\ \citenamefont
  {Prost}}]{juelicher1997modeling}%
  \BibitemOpen
  \bibfield  {author} {\bibinfo {author} {\bibfnamefont {F.}~\bibnamefont
  {J\"ulicher}}, \bibinfo {author} {\bibfnamefont {A.}~\bibnamefont {Ajdari}},\
  and\ \bibinfo {author} {\bibfnamefont {J.}~\bibnamefont {Prost}},\ }\bibfield
   {title} {\bibinfo {title} {Modeling molecular motors},\ }\href
  {https://doi.org/10.1103/RevModPhys.69.1269} {\bibfield  {journal} {\bibinfo
  {journal} {Rev. Mod. Phys.}\ }\textbf {\bibinfo {volume} {69}},\ \bibinfo
  {pages} {1269} (\bibinfo {year} {1997})}\BibitemShut {NoStop}%
\bibitem [{\citenamefont {Kolomeisky}\ and\ \citenamefont
  {Fisher}(2007)}]{Kolomeisky2007}%
  \BibitemOpen
  \bibfield  {author} {\bibinfo {author} {\bibfnamefont {A.~B.}\ \bibnamefont
  {Kolomeisky}}\ and\ \bibinfo {author} {\bibfnamefont {M.~E.}\ \bibnamefont
  {Fisher}},\ }\bibfield  {title} {\bibinfo {title} {Molecular motors: A
  theorist’s perspective},\ }\href
  {https://doi.org/10.1146/annurev.physchem.58.032806.104532} {\bibfield
  {journal} {\bibinfo  {journal} {Annu. Rev. Phys. Chem.}\ }\textbf {\bibinfo
  {volume} {58}},\ \bibinfo {pages} {675–695} (\bibinfo {year}
  {2007})}\BibitemShut {NoStop}%
\bibitem [{\citenamefont {Mugnai}\ \emph {et~al.}(2020)\citenamefont {Mugnai},
  \citenamefont {Hyeon}, \citenamefont {Hinczewski},\ and\ \citenamefont
  {Thirumalai}}]{RevModPhys.92.025001}%
  \BibitemOpen
  \bibfield  {author} {\bibinfo {author} {\bibfnamefont {M.~L.}\ \bibnamefont
  {Mugnai}}, \bibinfo {author} {\bibfnamefont {C.}~\bibnamefont {Hyeon}},
  \bibinfo {author} {\bibfnamefont {M.}~\bibnamefont {Hinczewski}},\ and\
  \bibinfo {author} {\bibfnamefont {D.}~\bibnamefont {Thirumalai}},\ }\bibfield
   {title} {\bibinfo {title} {Theoretical perspectives on biological
  machines},\ }\href {https://doi.org/10.1103/RevModPhys.92.025001} {\bibfield
  {journal} {\bibinfo  {journal} {Rev. Mod. Phys.}\ }\textbf {\bibinfo {volume}
  {92}},\ \bibinfo {pages} {025001} (\bibinfo {year} {2020})}\BibitemShut
  {NoStop}%
\bibitem [{\citenamefont {Shi}\ \emph {et~al.}(2022)\citenamefont {Shi},
  \citenamefont {Pumm}, \citenamefont {Isensee}, \citenamefont {Zhao},
  \citenamefont {Verschueren}, \citenamefont {Martin-Gonzalez}, \citenamefont
  {Golestanian}, \citenamefont {Dietz},\ and\ \citenamefont
  {Dekker}}]{Shi2022}%
  \BibitemOpen
  \bibfield  {author} {\bibinfo {author} {\bibfnamefont {X.}~\bibnamefont
  {Shi}}, \bibinfo {author} {\bibfnamefont {A.-K.}\ \bibnamefont {Pumm}},
  \bibinfo {author} {\bibfnamefont {J.}~\bibnamefont {Isensee}}, \bibinfo
  {author} {\bibfnamefont {W.}~\bibnamefont {Zhao}}, \bibinfo {author}
  {\bibfnamefont {D.}~\bibnamefont {Verschueren}}, \bibinfo {author}
  {\bibfnamefont {A.}~\bibnamefont {Martin-Gonzalez}}, \bibinfo {author}
  {\bibfnamefont {R.}~\bibnamefont {Golestanian}}, \bibinfo {author}
  {\bibfnamefont {H.}~\bibnamefont {Dietz}},\ and\ \bibinfo {author}
  {\bibfnamefont {C.}~\bibnamefont {Dekker}},\ }\bibfield  {title} {\bibinfo
  {title} {Sustained unidirectional rotation of a self-organized {DNA} rotor on
  a nanopore},\ }\href {https://doi.org/10.1038/s41567-022-01683-z} {\bibfield
  {journal} {\bibinfo  {journal} {Nat. Phys.}\ }\textbf {\bibinfo {volume}
  {18}},\ \bibinfo {pages} {1105–1111} (\bibinfo {year} {2022})}\BibitemShut
  {NoStop}%
\bibitem [{\citenamefont {Shi}\ \emph {et~al.}(2023)\citenamefont {Shi},
  \citenamefont {Pumm}, \citenamefont {Maffeo}, \citenamefont {Kohler},
  \citenamefont {Feigl}, \citenamefont {Zhao}, \citenamefont {Verschueren},
  \citenamefont {Golestanian}, \citenamefont {Aksimentiev}, \citenamefont
  {Dietz},\ and\ \citenamefont {Dekker}}]{Shi2023}%
  \BibitemOpen
  \bibfield  {author} {\bibinfo {author} {\bibfnamefont {X.}~\bibnamefont
  {Shi}}, \bibinfo {author} {\bibfnamefont {A.-K.}\ \bibnamefont {Pumm}},
  \bibinfo {author} {\bibfnamefont {C.}~\bibnamefont {Maffeo}}, \bibinfo
  {author} {\bibfnamefont {F.}~\bibnamefont {Kohler}}, \bibinfo {author}
  {\bibfnamefont {E.}~\bibnamefont {Feigl}}, \bibinfo {author} {\bibfnamefont
  {W.}~\bibnamefont {Zhao}}, \bibinfo {author} {\bibfnamefont {D.}~\bibnamefont
  {Verschueren}}, \bibinfo {author} {\bibfnamefont {R.}~\bibnamefont
  {Golestanian}}, \bibinfo {author} {\bibfnamefont {A.}~\bibnamefont
  {Aksimentiev}}, \bibinfo {author} {\bibfnamefont {H.}~\bibnamefont {Dietz}},\
  and\ \bibinfo {author} {\bibfnamefont {C.}~\bibnamefont {Dekker}},\
  }\bibfield  {title} {\bibinfo {title} {A {DNA} turbine powered by a
  transmembrane potential across a nanopore},\ }\href
  {http://dx.doi.org/10.1038/s41565-023-01527-8} {\bibfield  {journal}
  {\bibinfo  {journal} {Nat. Nanotechnol.}\ } (\bibinfo {year}
  {2023})}\BibitemShut {NoStop}%
\bibitem [{\citenamefont {Golestanian}(2010)}]{RG2010}%
  \BibitemOpen
  \bibfield  {author} {\bibinfo {author} {\bibfnamefont {R.}~\bibnamefont
  {Golestanian}},\ }\bibfield  {title} {\bibinfo {title} {Synthetic
  mechanochemical molecular swimmer},\ }\href
  {https://doi.org/10.1103/PhysRevLett.105.018103} {\bibfield  {journal}
  {\bibinfo  {journal} {Phys. Rev. Lett.}\ }\textbf {\bibinfo {volume} {105}},\
  \bibinfo {pages} {018103} (\bibinfo {year} {2010})}\BibitemShut {NoStop}%
\bibitem [{\citenamefont {Golestanian}\ and\ \citenamefont
  {Ajdari}(2008)}]{RG2008}%
  \BibitemOpen
  \bibfield  {author} {\bibinfo {author} {\bibfnamefont {R.}~\bibnamefont
  {Golestanian}}\ and\ \bibinfo {author} {\bibfnamefont {A.}~\bibnamefont
  {Ajdari}},\ }\bibfield  {title} {\bibinfo {title} {Mechanical response of a
  small swimmer driven by conformational transitions},\ }\href
  {https://doi.org/10.1103/PhysRevLett.100.038101} {\bibfield  {journal}
  {\bibinfo  {journal} {Phys. Rev. Lett.}\ }\textbf {\bibinfo {volume} {100}},\
  \bibinfo {pages} {038101} (\bibinfo {year} {2008})}\BibitemShut {NoStop}%
\bibitem [{\citenamefont {Golestanian}(2009)}]{RG2009stoch}%
  \BibitemOpen
  \bibfield  {author} {\bibinfo {author} {\bibfnamefont {R.}~\bibnamefont
  {Golestanian}},\ }\bibfield  {title} {\bibinfo {title} {Anomalous diffusion
  of symmetric and asymmetric active colloids},\ }\href
  {https://doi.org/10.1103/PhysRevLett.102.188305} {\bibfield  {journal}
  {\bibinfo  {journal} {Phys. Rev. Lett.}\ }\textbf {\bibinfo {volume} {102}},\
  \bibinfo {pages} {188305} (\bibinfo {year} {2009})}\BibitemShut {NoStop}%
\bibitem [{\citenamefont {Golestanian}(2015)}]{RG-prl2015}%
  \BibitemOpen
  \bibfield  {author} {\bibinfo {author} {\bibfnamefont {R.}~\bibnamefont
  {Golestanian}},\ }\bibfield  {title} {\bibinfo {title} {Enhanced diffusion of
  enzymes that catalyze exothermic reactions},\ }\href
  {https://doi.org/10.1103/PhysRevLett.115.108102} {\bibfield  {journal}
  {\bibinfo  {journal} {Phys. Rev. Lett.}\ }\textbf {\bibinfo {volume} {115}},\
  \bibinfo {pages} {108102} (\bibinfo {year} {2015})}\BibitemShut {NoStop}%
\bibitem [{\citenamefont {Illien}\ \emph {et~al.}(2017)\citenamefont {Illien},
  \citenamefont {Adeleke-Larodo},\ and\ \citenamefont
  {Golestanian}}]{Illien2017}%
  \BibitemOpen
  \bibfield  {author} {\bibinfo {author} {\bibfnamefont {P.}~\bibnamefont
  {Illien}}, \bibinfo {author} {\bibfnamefont {T.}~\bibnamefont
  {Adeleke-Larodo}},\ and\ \bibinfo {author} {\bibfnamefont {R.}~\bibnamefont
  {Golestanian}},\ }\bibfield  {title} {\bibinfo {title} {Diffusion of an
  enzyme: The role of fluctuation-induced hydrodynamic coupling},\ }\href
  {https://doi.org/10.1209/0295-5075/119/40002} {\bibfield  {journal} {\bibinfo
   {journal} {EPL}\ }\textbf {\bibinfo {volume} {119}},\ \bibinfo {pages}
  {40002} (\bibinfo {year} {2017})}\BibitemShut {NoStop}%
\bibitem [{\citenamefont {Agudo-Canalejo}\ \emph {et~al.}(2018)\citenamefont
  {Agudo-Canalejo}, \citenamefont {Adeleke-Larodo}, \citenamefont {Illien},\
  and\ \citenamefont {Golestanian}}]{AgudoCanalejo2018}%
  \BibitemOpen
  \bibfield  {author} {\bibinfo {author} {\bibfnamefont {J.}~\bibnamefont
  {Agudo-Canalejo}}, \bibinfo {author} {\bibfnamefont {T.}~\bibnamefont
  {Adeleke-Larodo}}, \bibinfo {author} {\bibfnamefont {P.}~\bibnamefont
  {Illien}},\ and\ \bibinfo {author} {\bibfnamefont {R.}~\bibnamefont
  {Golestanian}},\ }\bibfield  {title} {\bibinfo {title} {Enhanced diffusion
  and chemotaxis at the nanoscale},\ }\href
  {https://doi.org/10.1021/acs.accounts.8b00280} {\bibfield  {journal}
  {\bibinfo  {journal} {Acc. Chem. Res.}\ }\textbf {\bibinfo {volume} {51}},\
  \bibinfo {pages} {2365–2372} (\bibinfo {year} {2018})}\BibitemShut
  {NoStop}%
\bibitem [{\citenamefont {Agudo-Canalejo}\ \emph {et~al.}(2021)\citenamefont
  {Agudo-Canalejo}, \citenamefont {Adeleke-Larodo}, \citenamefont {Illien},\
  and\ \citenamefont {Golestanian}}]{jaime}%
  \BibitemOpen
  \bibfield  {author} {\bibinfo {author} {\bibfnamefont {J.}~\bibnamefont
  {Agudo-Canalejo}}, \bibinfo {author} {\bibfnamefont {T.}~\bibnamefont
  {Adeleke-Larodo}}, \bibinfo {author} {\bibfnamefont {P.}~\bibnamefont
  {Illien}},\ and\ \bibinfo {author} {\bibfnamefont {R.}~\bibnamefont
  {Golestanian}},\ }\bibfield  {title} {\bibinfo {title} {Synchronization and
  enhanced catalysis of mechanically coupled enzymes},\ }\href
  {https://doi.org/10.1103/PhysRevLett.127.208103} {\bibfield  {journal}
  {\bibinfo  {journal} {Phys. Rev. Lett.}\ }\textbf {\bibinfo {volume} {127}},\
  \bibinfo {pages} {208103} (\bibinfo {year} {2021})}\BibitemShut {NoStop}%
\bibitem [{\citenamefont {Zeravcic}\ and\ \citenamefont
  {Brenner}(2014)}]{zeravcic2014self}%
  \BibitemOpen
  \bibfield  {author} {\bibinfo {author} {\bibfnamefont {Z.}~\bibnamefont
  {Zeravcic}}\ and\ \bibinfo {author} {\bibfnamefont {M.~P.}\ \bibnamefont
  {Brenner}},\ }\bibfield  {title} {\bibinfo {title} {Self-replicating
  colloidal clusters},\ }\href {https://doi.org/10.1073/pnas.1313601111}
  {\bibfield  {journal} {\bibinfo  {journal} {Proc. Natl. Acad. Sci. U.S.A.}\
  }\textbf {\bibinfo {volume} {111}},\ \bibinfo {pages} {1748–1753} (\bibinfo
  {year} {2014})}\BibitemShut {NoStop}%
\bibitem [{\citenamefont {Zeravcic}\ and\ \citenamefont
  {Brenner}(2017)}]{zeravcic2017spontaneous}%
  \BibitemOpen
  \bibfield  {author} {\bibinfo {author} {\bibfnamefont {Z.}~\bibnamefont
  {Zeravcic}}\ and\ \bibinfo {author} {\bibfnamefont {M.~P.}\ \bibnamefont
  {Brenner}},\ }\bibfield  {title} {\bibinfo {title} {Spontaneous emergence of
  catalytic cycles with colloidal spheres},\ }\href
  {https://doi.org/10.1073/pnas.1611959114} {\bibfield  {journal} {\bibinfo
  {journal} {Proc. Natl. Acad. Sci. U.S.A.}\ }\textbf {\bibinfo {volume}
  {114}},\ \bibinfo {pages} {4342–4347} (\bibinfo {year} {2017})}\BibitemShut
  {NoStop}%
\bibitem [{\citenamefont {Zeravcic}\ \emph {et~al.}(2017)\citenamefont
  {Zeravcic}, \citenamefont {Manoharan},\ and\ \citenamefont
  {Brenner}}]{zeravcic2017colloquium}%
  \BibitemOpen
  \bibfield  {author} {\bibinfo {author} {\bibfnamefont {Z.}~\bibnamefont
  {Zeravcic}}, \bibinfo {author} {\bibfnamefont {V.~N.}\ \bibnamefont
  {Manoharan}},\ and\ \bibinfo {author} {\bibfnamefont {M.~P.}\ \bibnamefont
  {Brenner}},\ }\bibfield  {title} {\bibinfo {title} {Colloquium: Toward living
  matter with colloidal particles},\ }\href
  {https://doi.org/10.1103/RevModPhys.89.031001} {\bibfield  {journal}
  {\bibinfo  {journal} {Rev. Mod. Phys.}\ }\textbf {\bibinfo {volume} {89}},\
  \bibinfo {pages} {031001} (\bibinfo {year} {2017})}\BibitemShut {NoStop}%
\bibitem [{\citenamefont {Rivoire}(2020)}]{rivoire2020geometry}%
  \BibitemOpen
  \bibfield  {author} {\bibinfo {author} {\bibfnamefont {O.}~\bibnamefont
  {Rivoire}},\ }\bibfield  {title} {\bibinfo {title} {Geometry and flexibility
  of optimal catalysts in a minimal elastic model},\ }\href
  {https://doi.org/10.1021/acs.jpcb.0c00244} {\bibfield  {journal} {\bibinfo
  {journal} {J. Phys. Chem. B}\ }\textbf {\bibinfo {volume} {124}},\ \bibinfo
  {pages} {807–813} (\bibinfo {year} {2020})}\BibitemShut {NoStop}%
\bibitem [{\citenamefont {McMullen}\ \emph {et~al.}(2022)\citenamefont
  {McMullen}, \citenamefont {Muñoz~Basagoiti}, \citenamefont {Zeravcic},\ and\
  \citenamefont {Brujic}}]{McMullen2022}%
  \BibitemOpen
  \bibfield  {author} {\bibinfo {author} {\bibfnamefont {A.}~\bibnamefont
  {McMullen}}, \bibinfo {author} {\bibfnamefont {M.}~\bibnamefont
  {Muñoz~Basagoiti}}, \bibinfo {author} {\bibfnamefont {Z.}~\bibnamefont
  {Zeravcic}},\ and\ \bibinfo {author} {\bibfnamefont {J.}~\bibnamefont
  {Brujic}},\ }\bibfield  {title} {\bibinfo {title} {Self-assembly of emulsion
  droplets through programmable folding},\ }\href
  {https://doi.org/10.1038/s41586-022-05198-8} {\bibfield  {journal} {\bibinfo
  {journal} {Nature}\ }\textbf {\bibinfo {volume} {610}},\ \bibinfo {pages}
  {502–506} (\bibinfo {year} {2022})}\BibitemShut {NoStop}%
\bibitem [{\citenamefont {Mu{\~{n}}oz-Basagoiti}\ \emph
  {et~al.}(2023)\citenamefont {Mu{\~{n}}oz-Basagoiti}, \citenamefont
  {Rivoire},\ and\ \citenamefont {Zeravcic}}]{munoz2023computational}%
  \BibitemOpen
  \bibfield  {author} {\bibinfo {author} {\bibfnamefont {M.}~\bibnamefont
  {Mu{\~{n}}oz-Basagoiti}}, \bibinfo {author} {\bibfnamefont {O.}~\bibnamefont
  {Rivoire}},\ and\ \bibinfo {author} {\bibfnamefont {Z.}~\bibnamefont
  {Zeravcic}},\ }\bibfield  {title} {\bibinfo {title} {Computational design of
  a minimal catalyst using colloidal particles with programmable
  interactions},\ }\href {https://doi.org/10.1039/d3sm00194f} {\bibfield
  {journal} {\bibinfo  {journal} {Soft Matter}\ }\textbf {\bibinfo {volume}
  {19}},\ \bibinfo {pages} {3933} (\bibinfo {year} {2023})}\BibitemShut
  {NoStop}%
\bibitem [{\citenamefont {Rivoire}(2023)}]{rivoire2023flexibility}%
  \BibitemOpen
  \bibfield  {author} {\bibinfo {author} {\bibfnamefont {O.}~\bibnamefont
  {Rivoire}},\ }\bibfield  {title} {\bibinfo {title} {How flexibility can
  enhance catalysis},\ }\href {https://doi.org/10.1103/PhysRevLett.131.088401}
  {\bibfield  {journal} {\bibinfo  {journal} {Phys. Rev. Lett.}\ }\textbf
  {\bibinfo {volume} {131}},\ \bibinfo {pages} {088401} (\bibinfo {year}
  {2023})}\BibitemShut {NoStop}%
\bibitem [{\citenamefont {Tang}\ \emph {et~al.}(2025)\citenamefont {Tang},
  \citenamefont {Wu}, \citenamefont {Wu}, \citenamefont {Tang}, \citenamefont
  {Zhang}, \citenamefont {Zhu}, \citenamefont {Zhu},\ and\ \citenamefont
  {Chen}}]{Tang2025Jan}%
  \BibitemOpen
  \bibfield  {author} {\bibinfo {author} {\bibfnamefont {Z.}~\bibnamefont
  {Tang}}, \bibinfo {author} {\bibfnamefont {J.}~\bibnamefont {Wu}}, \bibinfo
  {author} {\bibfnamefont {S.}~\bibnamefont {Wu}}, \bibinfo {author}
  {\bibfnamefont {W.}~\bibnamefont {Tang}}, \bibinfo {author} {\bibfnamefont
  {J.-R.}\ \bibnamefont {Zhang}}, \bibinfo {author} {\bibfnamefont
  {W.}~\bibnamefont {Zhu}}, \bibinfo {author} {\bibfnamefont {J.-J.}\
  \bibnamefont {Zhu}},\ and\ \bibinfo {author} {\bibfnamefont {Z.}~\bibnamefont
  {Chen}},\ }\bibfield  {title} {\bibinfo {title} {{Single
  molecule{\textendash}driven nanomotors reveal the dynamic-disordered
  chemomechanical transduction of active enzymes}},\ }\href
  {https://www.science.org/doi/10.1126/sciadv.ads0446} {\bibfield  {journal}
  {\bibinfo  {journal} {Sci. Adv.}\ }\textbf {\bibinfo {volume} {11}} (\bibinfo
  {year} {2025})}\BibitemShut {NoStop}%
\bibitem [{\citenamefont {Alberts}\ \emph {et~al.}(1994)\citenamefont
  {Alberts}, \citenamefont {Bray}, \citenamefont {Lewis}, \citenamefont {Raff},
  \citenamefont {Roberts}, \citenamefont {Watson} \emph
  {et~al.}}]{alberts2017molecular}%
  \BibitemOpen
  \bibfield  {author} {\bibinfo {author} {\bibfnamefont {B.}~\bibnamefont
  {Alberts}}, \bibinfo {author} {\bibfnamefont {D.}~\bibnamefont {Bray}},
  \bibinfo {author} {\bibfnamefont {J.}~\bibnamefont {Lewis}}, \bibinfo
  {author} {\bibfnamefont {M.}~\bibnamefont {Raff}}, \bibinfo {author}
  {\bibfnamefont {K.}~\bibnamefont {Roberts}}, \bibinfo {author} {\bibfnamefont
  {J.~D.}\ \bibnamefont {Watson}}, \emph {et~al.},\ }\href@noop {} {\emph
  {\bibinfo {title} {Molecular biology of the cell}}},\ Vol.~\bibinfo {volume}
  {3}\ (\bibinfo  {publisher} {Garland New York},\ \bibinfo {year}
  {1994})\BibitemShut {NoStop}%
\bibitem [{\citenamefont {Borsley}\ \emph {et~al.}(2022)\citenamefont
  {Borsley}, \citenamefont {Leigh},\ and\ \citenamefont
  {Roberts}}]{Borsley2022}%
  \BibitemOpen
  \bibfield  {author} {\bibinfo {author} {\bibfnamefont {S.}~\bibnamefont
  {Borsley}}, \bibinfo {author} {\bibfnamefont {D.~A.}\ \bibnamefont {Leigh}},\
  and\ \bibinfo {author} {\bibfnamefont {B.~M.~W.}\ \bibnamefont {Roberts}},\
  }\bibfield  {title} {\bibinfo {title} {Chemical fuels for molecular
  machinery},\ }\href {https://doi.org/10.1038/s41557-022-00970-9} {\bibfield
  {journal} {\bibinfo  {journal} {Nat. Chem.}\ }\textbf {\bibinfo {volume}
  {14}},\ \bibinfo {pages} {728} (\bibinfo {year} {2022})}\BibitemShut
  {NoStop}%
\bibitem [{\citenamefont {Astumian}\ and\ \citenamefont
  {Bier}(1996)}]{Astumian1996}%
  \BibitemOpen
  \bibfield  {author} {\bibinfo {author} {\bibfnamefont {R.}~\bibnamefont
  {Astumian}}\ and\ \bibinfo {author} {\bibfnamefont {M.}~\bibnamefont
  {Bier}},\ }\bibfield  {title} {\bibinfo {title} {Mechanochemical coupling of
  the motion of molecular motors to {ATP} hydrolysis},\ }\href
  {https://doi.org/10.1016/s0006-3495(96)79605-4} {\bibfield  {journal}
  {\bibinfo  {journal} {Biophys. J.}\ }\textbf {\bibinfo {volume} {70}},\
  \bibinfo {pages} {637} (\bibinfo {year} {1996})}\BibitemShut {NoStop}%
\bibitem [{\citenamefont {Eide}\ \emph {et~al.}(2006)\citenamefont {Eide},
  \citenamefont {Chakraborty},\ and\ \citenamefont {Oster}}]{Eide2006}%
  \BibitemOpen
  \bibfield  {author} {\bibinfo {author} {\bibfnamefont {J.~L.}\ \bibnamefont
  {Eide}}, \bibinfo {author} {\bibfnamefont {A.~K.}\ \bibnamefont
  {Chakraborty}},\ and\ \bibinfo {author} {\bibfnamefont {G.~F.}\ \bibnamefont
  {Oster}},\ }\bibfield  {title} {\bibinfo {title} {Simple models for
  extracting mechanical work from the {ATP} hydrolysis cycle},\ }\href
  {https://doi.org/10.1529/biophysj.105.073320} {\bibfield  {journal} {\bibinfo
   {journal} {Biophys. J.}\ }\textbf {\bibinfo {volume} {90}},\ \bibinfo
  {pages} {4281} (\bibinfo {year} {2006})}\BibitemShut {NoStop}%
\bibitem [{\citenamefont {Chatzittofi}\ \emph
  {et~al.}(2023{\natexlab{a}})\citenamefont {Chatzittofi}, \citenamefont
  {Golestanian},\ and\ \citenamefont {Agudo-Canalejo}}]{chatzittofi}%
  \BibitemOpen
  \bibfield  {author} {\bibinfo {author} {\bibfnamefont {M.}~\bibnamefont
  {Chatzittofi}}, \bibinfo {author} {\bibfnamefont {R.}~\bibnamefont
  {Golestanian}},\ and\ \bibinfo {author} {\bibfnamefont {J.}~\bibnamefont
  {Agudo-Canalejo}},\ }\bibfield  {title} {\bibinfo {title} {Collective
  synchronization of dissipatively-coupled noise-activated processes},\ }\href
  {https://doi.org/10.1088/1367-2630/acf2bc} {\bibfield  {journal} {\bibinfo
  {journal} {New J. Phys.}\ }\textbf {\bibinfo {volume} {25}},\ \bibinfo
  {pages} {093014} (\bibinfo {year} {2023}{\natexlab{a}})}\BibitemShut
  {NoStop}%
\bibitem [{\citenamefont {Chatzittofi}\ \emph
  {et~al.}(2023{\natexlab{b}})\citenamefont {Chatzittofi}, \citenamefont
  {Golestanian},\ and\ \citenamefont
  {Agudo-Canalejo}}]{chatzittofi2023topological}%
  \BibitemOpen
  \bibfield  {author} {\bibinfo {author} {\bibfnamefont {M.}~\bibnamefont
  {Chatzittofi}}, \bibinfo {author} {\bibfnamefont {R.}~\bibnamefont
  {Golestanian}},\ and\ \bibinfo {author} {\bibfnamefont {J.}~\bibnamefont
  {Agudo-Canalejo}},\ }\bibfield  {title} {\bibinfo {title} {Topological phase
  locking in molecular oscillators},\ }\href
  {https://doi.org/10.48550/arXiv.2310.11788} {\bibfield  {journal} {\bibinfo
  {journal} {arXiv:2310.11788}\ } (\bibinfo {year}
  {2023}{\natexlab{b}})}\BibitemShut {NoStop}%
\bibitem [{\citenamefont {Lu}\ \emph {et~al.}(2014)\citenamefont {Lu},
  \citenamefont {Li},\ and\ \citenamefont {Zhang}}]{Lu2014}%
  \BibitemOpen
  \bibfield  {author} {\bibinfo {author} {\bibfnamefont {S.}~\bibnamefont
  {Lu}}, \bibinfo {author} {\bibfnamefont {S.}~\bibnamefont {Li}},\ and\
  \bibinfo {author} {\bibfnamefont {J.}~\bibnamefont {Zhang}},\ }\bibfield
  {title} {\bibinfo {title} {Harnessing allostery: A novel approach to drug
  discovery},\ }\href {https://doi.org/10.1002/med.21317} {\bibfield  {journal}
  {\bibinfo  {journal} {Med. Res. Rev.}\ }\textbf {\bibinfo {volume} {34}},\
  \bibinfo {pages} {1242} (\bibinfo {year} {2014})}\BibitemShut {NoStop}%
\bibitem [{\citenamefont {Hehir}\ \emph {et~al.}(2000)\citenamefont {Hehir},
  \citenamefont {Murphy},\ and\ \citenamefont {Kantrowitz}}]{Hehir2000}%
  \BibitemOpen
  \bibfield  {author} {\bibinfo {author} {\bibfnamefont {M.~J.}\ \bibnamefont
  {Hehir}}, \bibinfo {author} {\bibfnamefont {J.~E.}\ \bibnamefont {Murphy}},\
  and\ \bibinfo {author} {\bibfnamefont {E.~R.}\ \bibnamefont {Kantrowitz}},\
  }\bibfield  {title} {\bibinfo {title} {Characterization of heterodimeric
  alkaline phosphatases from escherichia coli: An investigation of intragenic
  complementation},\ }\href {https://doi.org/10.1006/jmbi.2000.4230} {\bibfield
   {journal} {\bibinfo  {journal} {J. Mol. Biol.}\ }\textbf {\bibinfo {volume}
  {304}},\ \bibinfo {pages} {645} (\bibinfo {year} {2000})}\BibitemShut
  {NoStop}%
\bibitem [{\citenamefont {De~Groot}\ and\ \citenamefont
  {Mazur}(2013)}]{degroot}%
  \BibitemOpen
  \bibfield  {author} {\bibinfo {author} {\bibfnamefont {S.~R.}\ \bibnamefont
  {De~Groot}}\ and\ \bibinfo {author} {\bibfnamefont {P.}~\bibnamefont
  {Mazur}},\ }\href@noop {} {\emph {\bibinfo {title} {Non-equilibrium
  thermodynamics}}}\ (\bibinfo  {publisher} {Courier Corporation},\ \bibinfo
  {year} {2013})\BibitemShut {NoStop}%
\bibitem [{\citenamefont {Kim}\ and\ \citenamefont
  {Karrila}(2013)}]{kim2013microhydrodynamics}%
  \BibitemOpen
  \bibfield  {author} {\bibinfo {author} {\bibfnamefont {S.}~\bibnamefont
  {Kim}}\ and\ \bibinfo {author} {\bibfnamefont {S.~J.}\ \bibnamefont
  {Karrila}},\ }\href@noop {} {\emph {\bibinfo {title} {Microhydrodynamics:
  principles and selected applications}}}\ (\bibinfo  {publisher} {Courier
  Corporation},\ \bibinfo {year} {2013})\BibitemShut {NoStop}%
\bibitem [{\citenamefont {Chatzittofi}\ \emph
  {et~al.}(2024{\natexlab{a}})\citenamefont {Chatzittofi}, \citenamefont
  {Agudo-Canalejo},\ and\ \citenamefont {Golestanian}}]{Chatzittofi2024Aug}%
  \BibitemOpen
  \bibfield  {author} {\bibinfo {author} {\bibfnamefont {M.}~\bibnamefont
  {Chatzittofi}}, \bibinfo {author} {\bibfnamefont {J.}~\bibnamefont
  {Agudo-Canalejo}},\ and\ \bibinfo {author} {\bibfnamefont {R.}~\bibnamefont
  {Golestanian}},\ }\bibfield  {title} {\bibinfo {title} {{Nonlinear response
  theory of molecular machines}},\ }\href
  {https://doi.org/10.1209/0295-5075/ad6a7e} {\bibfield  {journal} {\bibinfo
  {journal} {EPL}\ }\textbf {\bibinfo {volume} {147}},\ \bibinfo {pages}
  {21002} (\bibinfo {year} {2024}{\natexlab{a}})}\BibitemShut {NoStop}%
\bibitem [{\citenamefont {Risken}(1996)}]{risken1996fokker}%
  \BibitemOpen
  \bibfield  {author} {\bibinfo {author} {\bibfnamefont {H.}~\bibnamefont
  {Risken}},\ }\href@noop {} {\emph {\bibinfo {title} {{Fokker-Planck}
  equation}}}\ (\bibinfo  {publisher} {Springer},\ \bibinfo {year} {1996})\
  pp.\ \bibinfo {pages} {63--95}\BibitemShut {NoStop}%
\bibitem [{\citenamefont {Reimann}\ \emph {et~al.}(2001)\citenamefont
  {Reimann}, \citenamefont {Van~den Broeck}, \citenamefont {Linke},
  \citenamefont {H\"anggi}, \citenamefont {Rubi},\ and\ \citenamefont
  {P\'erez-Madrid}}]{PhysRevLett.87.010602}%
  \BibitemOpen
  \bibfield  {author} {\bibinfo {author} {\bibfnamefont {P.}~\bibnamefont
  {Reimann}}, \bibinfo {author} {\bibfnamefont {C.}~\bibnamefont {Van~den
  Broeck}}, \bibinfo {author} {\bibfnamefont {H.}~\bibnamefont {Linke}},
  \bibinfo {author} {\bibfnamefont {P.}~\bibnamefont {H\"anggi}}, \bibinfo
  {author} {\bibfnamefont {J.~M.}\ \bibnamefont {Rubi}},\ and\ \bibinfo
  {author} {\bibfnamefont {A.}~\bibnamefont {P\'erez-Madrid}},\ }\bibfield
  {title} {\bibinfo {title} {Giant acceleration of free diffusion by use of
  tilted periodic potentials},\ }\href
  {https://doi.org/10.1103/PhysRevLett.87.010602} {\bibfield  {journal}
  {\bibinfo  {journal} {Phys. Rev. Lett.}\ }\textbf {\bibinfo {volume} {87}},\
  \bibinfo {pages} {010602} (\bibinfo {year} {2001})}\BibitemShut {NoStop}%
\bibitem [{\citenamefont {Reimann}\ \emph {et~al.}(2002)\citenamefont
  {Reimann}, \citenamefont {Van~den Broeck}, \citenamefont {Linke},
  \citenamefont {H\"anggi}, \citenamefont {Rubi},\ and\ \citenamefont
  {P\'erez-Madrid}}]{PhysRevE.65.031104}%
  \BibitemOpen
  \bibfield  {author} {\bibinfo {author} {\bibfnamefont {P.}~\bibnamefont
  {Reimann}}, \bibinfo {author} {\bibfnamefont {C.}~\bibnamefont {Van~den
  Broeck}}, \bibinfo {author} {\bibfnamefont {H.}~\bibnamefont {Linke}},
  \bibinfo {author} {\bibfnamefont {P.}~\bibnamefont {H\"anggi}}, \bibinfo
  {author} {\bibfnamefont {J.~M.}\ \bibnamefont {Rubi}},\ and\ \bibinfo
  {author} {\bibfnamefont {A.}~\bibnamefont {P\'erez-Madrid}},\ }\bibfield
  {title} {\bibinfo {title} {Diffusion in tilted periodic potentials:
  Enhancement, universality, and scaling},\ }\href
  {https://doi.org/10.1103/PhysRevE.65.031104} {\bibfield  {journal} {\bibinfo
  {journal} {Phys. Rev. E}\ }\textbf {\bibinfo {volume} {65}},\ \bibinfo
  {pages} {031104} (\bibinfo {year} {2002})}\BibitemShut {NoStop}%
\bibitem [{\citenamefont {Langer}(1968)}]{langer1968theory}%
  \BibitemOpen
  \bibfield  {author} {\bibinfo {author} {\bibfnamefont {J.~S.}\ \bibnamefont
  {Langer}},\ }\bibfield  {title} {\bibinfo {title} {Theory of nucleation
  rates},\ }\href {https://doi.org/10.1103/PhysRevLett.21.973} {\bibfield
  {journal} {\bibinfo  {journal} {Phys. Rev. Lett.}\ }\textbf {\bibinfo
  {volume} {21}},\ \bibinfo {pages} {973} (\bibinfo {year} {1968})}\BibitemShut
  {NoStop}%
\bibitem [{\citenamefont {Langer}(1969)}]{langer1969statistical}%
  \BibitemOpen
  \bibfield  {author} {\bibinfo {author} {\bibfnamefont {J.}~\bibnamefont
  {Langer}},\ }\bibfield  {title} {\bibinfo {title} {Statistical theory of the
  decay of metastable states},\ }\href
  {https://doi.org/10.1016/0003-4916(69)90153-5} {\bibfield  {journal}
  {\bibinfo  {journal} {Ann. Phys.}\ }\textbf {\bibinfo {volume} {54}},\
  \bibinfo {pages} {258} (\bibinfo {year} {1969})}\BibitemShut {NoStop}%
\bibitem [{\citenamefont {Gutteridge}\ and\ \citenamefont
  {Thornton}(2005)}]{Gutteridge2005Feb}%
  \BibitemOpen
  \bibfield  {author} {\bibinfo {author} {\bibfnamefont {A.}~\bibnamefont
  {Gutteridge}}\ and\ \bibinfo {author} {\bibfnamefont {J.}~\bibnamefont
  {Thornton}},\ }\bibfield  {title} {\bibinfo {title} {{Conformational Changes
  Observed in Enzyme Crystal Structures upon Substrate Binding}},\ }\href
  {https://doi.org/10.1016/j.jmb.2004.11.013} {\bibfield  {journal} {\bibinfo
  {journal} {J. Mol. Biol.}\ }\textbf {\bibinfo {volume} {346}},\ \bibinfo
  {pages} {21} (\bibinfo {year} {2005})}\BibitemShut {NoStop}%
\bibitem [{\citenamefont {Lu}(2014)}]{Lu2014Jan}%
  \BibitemOpen
  \bibfield  {author} {\bibinfo {author} {\bibfnamefont {H.~P.}\ \bibnamefont
  {Lu}},\ }\bibfield  {title} {\bibinfo {title} {{Sizing up single-molecule
  enzymatic conformational dynamics}},\ }\href
  {https://doi.org/10.1039/C3CS60191A} {\bibfield  {journal} {\bibinfo
  {journal} {Chem. Soc. Rev.}\ }\textbf {\bibinfo {volume} {43}},\ \bibinfo
  {pages} {1118} (\bibinfo {year} {2014})}\BibitemShut {NoStop}%
\bibitem [{\citenamefont {Chong}\ \emph {et~al.}(2024)\citenamefont {Chong},
  \citenamefont {Oshima},\ and\ \citenamefont {Sugita}}]{Chong2024Nov}%
  \BibitemOpen
  \bibfield  {author} {\bibinfo {author} {\bibfnamefont {S.-H.}\ \bibnamefont
  {Chong}}, \bibinfo {author} {\bibfnamefont {H.}~\bibnamefont {Oshima}},\ and\
  \bibinfo {author} {\bibfnamefont {Y.}~\bibnamefont {Sugita}},\ }\bibfield
  {title} {\bibinfo {title} {{Allosteric changes in the conformational
  landscape of Src kinase upon substrate binding}},\ }\href
  {https://doi.org/10.1016/j.jmb.2024.168871} {\bibfield  {journal} {\bibinfo
  {journal} {J. Mol. Biol.}\ ,\ \bibinfo {pages} {168871}} (\bibinfo {year}
  {2024})}\BibitemShut {NoStop}%
\bibitem [{\citenamefont {Bennett}\ and\ \citenamefont
  {Steitz}(1978)}]{Bennett1978Oct}%
  \BibitemOpen
  \bibfield  {author} {\bibinfo {author} {\bibfnamefont {W.~S.}\ \bibnamefont
  {Bennett}}\ and\ \bibinfo {author} {\bibfnamefont {T.~A.}\ \bibnamefont
  {Steitz}},\ }\bibfield  {title} {\bibinfo {title} {{Glucose-induced
  conformational change in yeast hexokinase.}},\ }\href
  {https://doi.org/10.1073/pnas.75.10.4848} {\bibfield  {journal} {\bibinfo
  {journal} {Proc. Natl. Acad. Sci. U.S.A.}\ }\textbf {\bibinfo {volume}
  {75}},\ \bibinfo {pages} {4848} (\bibinfo {year} {1978})}\BibitemShut
  {NoStop}%
\bibitem [{\citenamefont {Kuser}\ \emph {et~al.}(2008)\citenamefont {Kuser},
  \citenamefont {Cupri}, \citenamefont {Bleicher},\ and\ \citenamefont
  {Polikarpov}}]{Kuser2008Aug}%
  \BibitemOpen
  \bibfield  {author} {\bibinfo {author} {\bibfnamefont {P.}~\bibnamefont
  {Kuser}}, \bibinfo {author} {\bibfnamefont {F.}~\bibnamefont {Cupri}},
  \bibinfo {author} {\bibfnamefont {L.}~\bibnamefont {Bleicher}},\ and\
  \bibinfo {author} {\bibfnamefont {I.}~\bibnamefont {Polikarpov}},\ }\bibfield
   {title} {\bibinfo {title} {{Crystal structure of yeast hexokinase PI in
  complex with glucose: A classical {\textquotedblleft}induced
  fit{\textquotedblright} example revised}},\ }\href
  {https://doi.org/10.1002/prot.21956} {\bibfield  {journal} {\bibinfo
  {journal} {Proteins Struct. Funct. Bioinf.}\ }\textbf {\bibinfo {volume}
  {72}},\ \bibinfo {pages} {731} (\bibinfo {year} {2008})}\BibitemShut
  {NoStop}%
\bibitem [{\citenamefont {Liang}\ \emph {et~al.}(2008)\citenamefont {Liang},
  \citenamefont {Woodward},\ and\ \citenamefont {Edelsbrunner}}]{Liang2008Dec}%
  \BibitemOpen
  \bibfield  {author} {\bibinfo {author} {\bibfnamefont {J.}~\bibnamefont
  {Liang}}, \bibinfo {author} {\bibfnamefont {C.}~\bibnamefont {Woodward}},\
  and\ \bibinfo {author} {\bibfnamefont {H.}~\bibnamefont {Edelsbrunner}},\
  }\bibfield  {title} {\bibinfo {title} {{Anatomy of protein pockets and
  cavities: Measurement of binding site geometry and implications for ligand
  design}},\ }\href {https://doi.org/10.1002/pro.5560070905} {\bibfield
  {journal} {\bibinfo  {journal} {Protein Sci.}\ }\textbf {\bibinfo {volume}
  {7}},\ \bibinfo {pages} {1884} (\bibinfo {year} {2008})}\BibitemShut
  {NoStop}%
\bibitem [{\citenamefont {Berg}\ \emph {et~al.}(2023)\citenamefont {Berg},
  \citenamefont {Gatto}, \citenamefont {Hines}, \citenamefont {Tymoczko},\ and\
  \citenamefont {Stryer}}]{stryer}%
  \BibitemOpen
  \bibfield  {author} {\bibinfo {author} {\bibfnamefont {J.}~\bibnamefont
  {Berg}}, \bibinfo {author} {\bibfnamefont {G.}~\bibnamefont {Gatto}},
  \bibinfo {author} {\bibfnamefont {J.}~\bibnamefont {Hines}}, \bibinfo
  {author} {\bibfnamefont {J.}~\bibnamefont {Tymoczko}},\ and\ \bibinfo
  {author} {\bibfnamefont {L.}~\bibnamefont {Stryer}},\ }\href@noop {} {\emph
  {\bibinfo {title} {Biochemistry}}}\ (\bibinfo  {publisher} {Macmillan
  Learning},\ \bibinfo {year} {2023})\BibitemShut {NoStop}%
\bibitem [{\citenamefont {Ragazzon}\ and\ \citenamefont
  {Prins}(2018)}]{Ragazzon2018}%
  \BibitemOpen
  \bibfield  {author} {\bibinfo {author} {\bibfnamefont {G.}~\bibnamefont
  {Ragazzon}}\ and\ \bibinfo {author} {\bibfnamefont {L.~J.}\ \bibnamefont
  {Prins}},\ }\bibfield  {title} {\bibinfo {title} {Energy consumption in
  chemical fuel-driven self-assembly},\ }\href
  {https://doi.org/10.1038/s41565-018-0250-8} {\bibfield  {journal} {\bibinfo
  {journal} {Nat. Nanotechnol.}\ }\textbf {\bibinfo {volume} {13}},\ \bibinfo
  {pages} {882–889} (\bibinfo {year} {2018})}\BibitemShut {NoStop}%
\bibitem [{\citenamefont {Chatzittofi}\ \emph
  {et~al.}(2024{\natexlab{b}})\citenamefont {Chatzittofi}, \citenamefont
  {Agudo-Canalejo},\ and\ \citenamefont
  {Golestanian}}]{chatzittofi2023entropy}%
  \BibitemOpen
  \bibfield  {author} {\bibinfo {author} {\bibfnamefont {M.}~\bibnamefont
  {Chatzittofi}}, \bibinfo {author} {\bibfnamefont {J.}~\bibnamefont
  {Agudo-Canalejo}},\ and\ \bibinfo {author} {\bibfnamefont {R.}~\bibnamefont
  {Golestanian}},\ }\bibfield  {title} {\bibinfo {title} {Entropy production
  and thermodynamic inference for stochastic microswimmers},\ }\href
  {https://doi.org/10.1103/PhysRevResearch.6.L022044} {\bibfield  {journal}
  {\bibinfo  {journal} {Phys. Rev. Res.}\ }\textbf {\bibinfo {volume} {6}},\
  \bibinfo {pages} {L022044} (\bibinfo {year}
  {2024}{\natexlab{b}})}\BibitemShut {NoStop}%
\bibitem [{\citenamefont {Frank}\ \emph {et~al.}(2024)\citenamefont {Frank},
  \citenamefont {Khoshouei}, \citenamefont {Fuss}, \citenamefont {Schiwietz},
  \citenamefont {Putz}, \citenamefont {Weber}, \citenamefont {Zhao},
  \citenamefont {Hattori}, \citenamefont {Feng}, \citenamefont {de~Stigter},
  \citenamefont {Ovchinnikov},\ and\ \citenamefont {Dietz}}]{Frank2024Oct}%
  \BibitemOpen
  \bibfield  {author} {\bibinfo {author} {\bibfnamefont {C.}~\bibnamefont
  {Frank}}, \bibinfo {author} {\bibfnamefont {A.}~\bibnamefont {Khoshouei}},
  \bibinfo {author} {\bibfnamefont {L.}~\bibnamefont {Fuss}}, \bibinfo {author}
  {\bibfnamefont {D.}~\bibnamefont {Schiwietz}}, \bibinfo {author}
  {\bibfnamefont {D.}~\bibnamefont {Putz}}, \bibinfo {author} {\bibfnamefont
  {L.}~\bibnamefont {Weber}}, \bibinfo {author} {\bibfnamefont
  {Z.}~\bibnamefont {Zhao}}, \bibinfo {author} {\bibfnamefont {M.}~\bibnamefont
  {Hattori}}, \bibinfo {author} {\bibfnamefont {S.}~\bibnamefont {Feng}},
  \bibinfo {author} {\bibfnamefont {Y.}~\bibnamefont {de~Stigter}}, \bibinfo
  {author} {\bibfnamefont {S.}~\bibnamefont {Ovchinnikov}},\ and\ \bibinfo
  {author} {\bibfnamefont {H.}~\bibnamefont {Dietz}},\ }\bibfield  {title}
  {\bibinfo {title} {{Scalable protein design using optimization in a relaxed
  sequence space}},\ }\href {https://doi.org/10.1126/science.adq1741}
  {\bibfield  {journal} {\bibinfo  {journal} {Science}\ }\textbf {\bibinfo
  {volume} {386}},\ \bibinfo {pages} {439} (\bibinfo {year}
  {2024})}\BibitemShut {NoStop}%
\bibitem [{\citenamefont {Inc.}(2023)}]{MATLAB}%
  \BibitemOpen
  \bibfield  {author} {\bibinfo {author} {\bibfnamefont {T.~M.}\ \bibnamefont
  {Inc.}},\ }\href {https://www.mathworks.com} {\bibinfo {title} {Matlab
  version: 9.14.0 (r2023a)}} (\bibinfo {year} {2023})\BibitemShut {NoStop}%
\bibitem [{\citenamefont {Van~Rossum}\ and\ \citenamefont
  {Drake}(2009)}]{python}%
  \BibitemOpen
  \bibfield  {author} {\bibinfo {author} {\bibfnamefont {G.}~\bibnamefont
  {Van~Rossum}}\ and\ \bibinfo {author} {\bibfnamefont {F.~L.}\ \bibnamefont
  {Drake}},\ }\href@noop {} {\emph {\bibinfo {title} {Python 3 Reference
  Manual}}}\ (\bibinfo  {publisher} {CreateSpace},\ \bibinfo {address} {Scotts
  Valley, CA},\ \bibinfo {year} {2009})\BibitemShut {NoStop}%
\bibitem [{\citenamefont {Bezanson}\ \emph {et~al.}(2012)\citenamefont
  {Bezanson}, \citenamefont {Karpinski}, \citenamefont {Shah},\ and\
  \citenamefont {Edelman}}]{julia}%
  \BibitemOpen
  \bibfield  {author} {\bibinfo {author} {\bibfnamefont {J.}~\bibnamefont
  {Bezanson}}, \bibinfo {author} {\bibfnamefont {S.}~\bibnamefont {Karpinski}},
  \bibinfo {author} {\bibfnamefont {V.~B.}\ \bibnamefont {Shah}},\ and\
  \bibinfo {author} {\bibfnamefont {A.}~\bibnamefont {Edelman}},\ }\bibfield
  {title} {\bibinfo {title} {Julia: A fast dynamic language for technical
  computing},\ }\href {https://doi.org/10.48550/arXiv.1209.5145} {\bibfield
  {journal} {\bibinfo  {journal} {arXiv:1209.5145}\ } (\bibinfo {year}
  {2012})}\BibitemShut {NoStop}%
\end{thebibliography}%

\begin{widetext}
\renewcommand{\thefigure}{S\arabic{figure}}
\setcounter{figure}{0}

\newpage

\setcounter{equation}{0}
\setcounter{figure}{0}
\setcounter{table}{0}
\renewcommand{\theequation}{S\arabic{equation}}
\renewcommand{\thefigure}{S\arabic{figure}}
\renewcommand{\thetable}{S\Roman{table}}

\begin{center}
\begin{title}\centering
\LARGE{\textit{{Supplemental} Information}}
\end{title}
\end{center}

\author{Michalis Chatzittofi}
\address{Max Planck Institute for Dynamics and Self-Organization (MPI-DS), D-37077 Göttingen, Germany} %Department of Living Matter Physics,
\author{Ramin Golestanian}
\email{ramin.golestanian@ds.mpg.de}
\address{Max Planck Institute for Dynamics and Self-Organization (MPI-DS), D-37077 Göttingen, Germany} %Department of Living Matter Physics, 
\address{Rudolf Peierls Centre for Theoretical Physics, University of Oxford, Oxford OX1 3PU, United Kingdom}
\author{Jaime Agudo-Canalejo}
\email{j.agudo-canalejo@ucl.ac.uk}
\address{Max Planck Institute for Dynamics and Self-Organization (MPI-DS), D-37077 Göttingen, Germany} %Department of Living Matter Physics, 
\address{Department of Physics and Astronomy, University College London, London WC1E 6BT, United Kingdom}

\maketitle
\date{\today}

%\tableofcontents

%\renewcommand{\thefigure}{S\arabic{figure}}
\renewcommand{\figurename}{Supplementary Fig.}
\setcounter{figure}{0}

%\newpage

\section*{Supplemental Methods}
\subsection*{Derivation of the phase-equations}\label{app:A}

We consider an enzyme and a molecule, both assumed to be dumbbell-shaped for simplicity. Upon binding, they are considered to be tightly linked together, thus forming a complex with three sub-units. The position vectors of the three sub-units are $\mathbf{r}_\ee$, $\mathbf{r}_\bb$ and $\mathbf{r}_\mm$ for the enzyme-only sub-unit, the bound enzyme-substrate sub-unit, and the molecule-only sub-unit, respectively. The lengths of the enzyme and molecule are therefore $\mathbf{L}_\ee = \mathbf{r}_\bb-\mathbf{r}_\ee = L_\ee \mathbf{\hat n}$ and $\mathbf{L}_\mm = \mathbf{r}_\mm-\mathbf{r}_\bb = L_\mm \mathbf{\hat n}$ where $\mathbf{\hat n}$ is the unit vector along the axis of the complex. The equations of motion for the sub-units are $\dot r_\ee = -\mu_\ee f_\ee$, $\dot r_\bb = \mu_\bb (f_\ee - f_\mm)$ and $\dot r_\mm = \mu_\mm f_\mm$, where $f_\ee$ and $f_\mm$ are the internal mechanical forces (stresses) of the enzyme and molecule, respectively, which can be expressed as $f_\ee= -\partial_{L_\ee}U(L_\ee,\phi_\ee)$ and $f_\mm = -\partial_{L_\mm}V_\mm(L_\mm)$. This implies that the equations of motion for the lengths are $\dot L_\ee = \mu_1 f_\ee - \mu_\bb f_\mm$ and $\dot L_\mm = -\mu_\bb f_\ee + \mu_2 f_\mm$, with the mobilities given as $\mu_1 \equiv \mu_\ee+\mu_\bb$ and $\mu_2 \equiv \mu_\mm + \mu_\bb$.

The governing dynamical equations for the two effective geometric degrees of freedom are therefore given by
\begin{align}
    \dot L_\ee = -\mu_1 k [L_\ee - L(\phi_\ee)] + \mu_\bb \partial_{L_\mm}V_\mm(L_\mm), \label{eq:L1}\\
    \dot L_\mm = \mu_\bb k[L_\ee - L(\phi_\ee)]-\mu_2 \partial_{L_\mm}V_\mm(L_\mm). \label{eq:L2}
\end{align}
The dynamics of the internal enzyme phase is in turn given by
\begin{align}\label{eq:phi1}
    \dot \phi_\ee &= \mu_\phi[ -\partial_{\phi_\ee} U(L_\ee,\phi_\ee)],\\ \nonumber 
    &= -\mu_\phi \left[-k(L_\ee - L(\phi_\ee))L'(\phi_\ee) + V'_\ee(\phi_\ee) \right],
\end{align}
with the prime representing a derivative with respect to $\phi_\ee$. We define $\delta L_\ee \equiv L_\ee - L(\phi_\ee)$ such that $\dot L_\ee = \dot{\delta L_\ee} + L'(\phi_\ee) \dot \phi_\ee$. If the enzyme is relatively stiff with the timescale $(\mu_1 k)^{-1}$ being much shorter than the timescale of changes in the internal phase, the enzyme length will quickly adapt to changes of the preferred length, and thus $\dot{\delta L_\ee} \approx 0$. From Equation~\eqref{eq:L1}, this implies
\begin{align}
    L'(\phi_\ee) \dot \phi_\ee \approx -\mu_1 k [L_\ee - L(\phi_\ee)] + \mu_\bb \partial_{L_\mm} V(L_\mm),
\end{align} 
or equivalently
\begin{align}
    k [L_\ee - L(\phi_\ee)] \approx \frac{1}{\mu_1}\left(\mu_\bb \partial_{L_\mm} V(L_\mm)-L'(\phi_\ee) \dot \phi_\ee\right). \label{eq:kLe}
\end{align}
Substitution of Equation~\eqref{eq:kLe} into Equations~ \eqref{eq:L2} and \eqref{eq:phi1} allows us to eliminate $k [L_\ee - L(\phi_\ee)]$ from these equations, which we then solve for $\dot \phi_\ee$ and $\dot L_\mm$. In this way, we project the dynamics onto the slow manifold $(\phi_\ee,L_\mm)$. Finally, we reexpress $L_\mm$ in terms of the dimensionless reaction coordinate $\phi_\mm$ through $L_\mm = L_\mm^{(0)} +\ell_\mm \phi_\mm$. The resulting governing equations for $(\phi_\ee,\phi_\mm)$ then correspond to Equation~(1) in the main article.

\subsection*{Reverse switch reaction}\label{app:B}

One can observe that for the switch reaction our enzyme is equally capable of catalyzing a reverse reaction, i.e. from long to short conformational state, by choosing $\varepsilon <0$. Exploiting the symmetries of our dynamical equations, we find that a $\pi$ phase shift will be needed for $\phi_\ee$ in the conformation cycle (Equation~8), as in this case the enzyme will first contract and then expand during its conformational change. A phase portrait of the deterministic dynamics at strong coupling (beyond the global bifurcation) is shown in Figure~\ref{fig:mirror}(A). This can be compared to the dynamics of the forward switch reaction in Figure~3(D). Note that the bifurcation now happens near the upper saddle point, where the blue solid line represents the flow from the long state to the short state. An example of the NESS distributions for this reverse case is shown in Figure~\ref{fig:mirror}(B), which may be compared to the case of the forward reaction in Figure~5(B). It highlights how the action of the enzyme can favour the short state even if the long state is thermodynamically preferred.

\subsection*{Steady state distribution for an uncoupled enzyme}\label{app:D}

In the uncoupled case with $h=0$ or $\ell_\ee/\ell_\mm =0$, an exact expression for the steady state probability distribution for $\phi_\ee$ can be found. In the absence of coupling, the dynamics of $\phi_\ee$ at the steady state is determined by solving the following equation for a constant flux
\begin{align}
    J_\mathrm{e} = - M_{\mathrm{ee}}(\phi_\ee)[V'_\ee(\phi_\ee) P_{\mathrm{e}} + k_{\rm B} T \partial_{\phi_\ee}P_{\mathrm{e}}].
\end{align}
Integrating this equation, we obtain
\begin{align}
    P_{\mathrm{e}} = e^{-V_\ee(\phi_\ee)/k_{\rm B} T} \left[ C - J_\mathrm{e} \int^{\phi_\ee}_0 M_{\mathrm{ee}}^{-1}(x)e^{V_\ee(x)/k_{\rm B} T} dx\right],
\end{align}
where $C$ and $J_\mathrm{e}$ are constants of integration and are determined by the boundary conditions. Note that $M_{\mathrm{ee}}(\phi_\ee)$ and $V_\ee'(\phi_\ee)$ are both $2\pi$-periodic. This implies that the probability distribution is also periodic, by following a similar derivation as in Ref.~\cite{risken1996fokker}. One finds that $C = \mathcal{N} I$ and $ (1- e^{-2\pi F/k_{\rm B}T}) \mathcal{N} = J_\mathrm{e}$, with $\mathcal{N}$ a normalization constant and
\begin{align}
    I = \int^{2\pi}_0 d x \, M_{\mathrm{ee}}^{-1}(x) \, e^{V_\ee(x)/k_{\rm B}T}.
\end{align}
Hence, the full expression for the steady state marginal probability distribution for the enzyme reaction coordinate in the uncoupled case is
\begin{align}\label{eq:risken}
    &P_{\mathrm{e}}(\phi_\ee) = \mathcal{N}\bigg[\int_{\phi_\ee}^{2\pi} d x \, M_{\mathrm{\ee \ee}}^{-1}(x)\exp\bigg(\frac{V_\ee(x)-V_\ee(\phi_\ee)}{k_{\rm B}T}\bigg) \nonumber\\
    &+ \int^{\phi_\ee}_0 d x \, M_{\ee \ee}^{-1}(x)\exp\bigg(\frac{V_\ee(x)-V_\ee(\phi_\ee)-2\pi F}{k_{\rm B}T}\bigg)\bigg].
\end{align}
The form of $V_\ee$ does not readily lend itself to further analytical progress, and we integrate Equation~\eqref{eq:risken} numerically to obtain the black lines in Figure~5(A,B). In the particular case $E_*=0$, we find that $J_\ee=0$ and thus the enzyme coordinate is in thermodynamic equilibrium described by the Boltzmann  weight $P_\ee(\phi_\ee) \propto e^{-V_\ee(\phi_\ee)/k_{\rm B} T}$.

\subsection*{Langer rate}\label{app:E}

Starting from the Fokker-Planck equation, one derives the Langer rate through the enzyme saddle point to be \cite{hanggi1990reaction,langer1968theory,langer1969statistical}
\begin{align}\label{eq:app_langers}
    k(\mathrm{fs} \to \mathrm{wp}) = \frac{\vert\Lambda_-\vert}{2 \pi} \frac{\sqrt{\lambda_\mathrm{e}^{(0)}\lambda_\mathrm{m}^{(0)}}}{\sqrt{\vert\lambda_\mathrm{e}^{(1)}\vert\lambda_\mathrm{m}^{(1)}}} \exp\bigg({-\frac{E_\mathrm{ba}}{k_{\rm B} T} }\bigg),
\end{align}
where $\lambda_\alpha^{(0)}=V''_\alpha(\phi_\alpha)|_{\rm min}$, and $\lambda_\alpha^{(1)}=V''_\alpha(\phi_\alpha)|_{\rm saddle}$, namely, the corresponding values evaluated around the saddle point. The definitions imply that $\lambda_\mathrm{m}^{(1)} = \lambda_\mathrm{m}^{(0)}$, and thus, they cancel each other in Equation \eqref{eq:app_langers}. Finally, $\Lambda_-$ is the negative eigenvalue of the matrix $\lambda_\alpha^{(1)} M_{\alpha \beta}$ evaluated around the saddle point which is given by
\begin{align}
  &  \Lambda_- = \frac{1}{2}\bigg( M_{\ee \ee} \lambda_\ee^{(1)} + M_{\mm \mm}\lambda_\mm^{(1)} \nonumber \\
  & \hskip5mm- \sqrt{(M_{\ee \ee} \lambda_\mathrm{e}^{(1)} - M_{\mm \mm}\lambda_\mm^{(1)})^2 + 4 M_{\ee \mm}^2\lambda_\ee^{(1)} \lambda_\mm^{(1)}} \bigg).
\end{align}

\section*{Supplemental Figures}
\bigskip
\begin{figure}[ht]
\centering
\includegraphics[width=0.9 \linewidth]{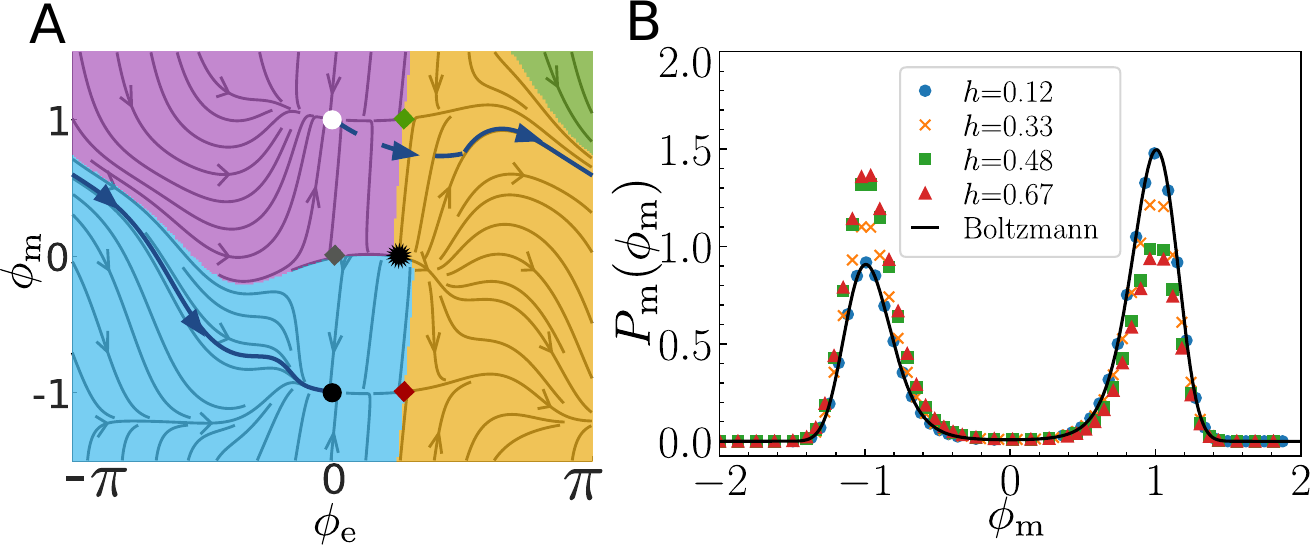}
\caption{\textbf{Reverse switch reaction.} (A) Phase-portrait for $h=0.48$, $\ell_\ee/\ell_\mm = 2.5$, and $\varepsilon/\Delta =-0.1$. The black solid line represents a deterministic trajectory. (B) Marginal distributions of the molecule reaction coordinate in steady-state, for various values of the coupling strength. (A) and (B) can be respectively compared to Figure~3(D) and Figure~5(C) for the forward reaction.}\label{fig:mirror}
\end{figure}

\begin{figure}[ht]
\centering
\includegraphics[width=0.9\linewidth]{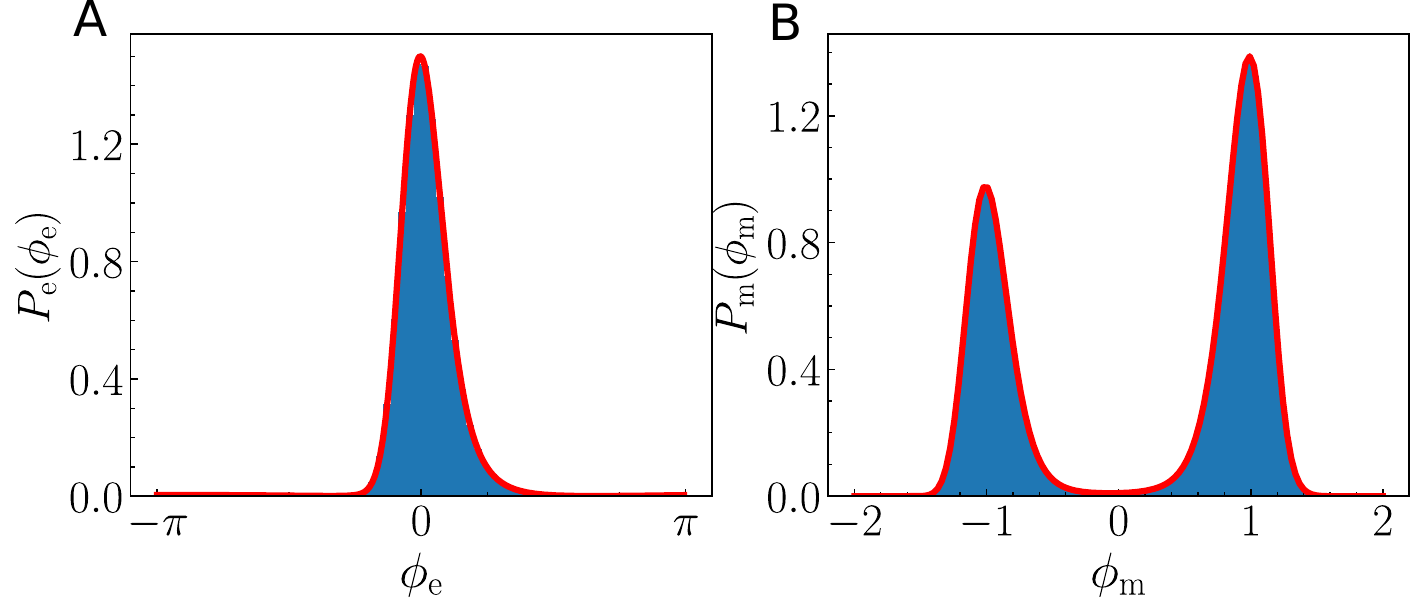}
\caption{\textbf{Comparison of the marginal distributions obtained from the Fokker-Planck equation at steady state and long simulations of the Langevin equations.} (A) and (B) show the distribution of the enzyme and molecule reaction coordinates, respectively. The parameters for this example are $h=0.67$ and $k_{\rm B}T/\Delta =0.2$. The red solid line is the solution of the Fokker-Planck equation and the histograms are obtained by the Langevin equations. }\label{fig:histogram_appendix}
\end{figure}

\end{widetext}

\end{document}